\documentclass[manuscript]{acmart}
\usepackage{amsmath,amsfonts}
\usepackage{array}
\usepackage{textcomp}
\usepackage{stfloats}
\usepackage{url}
\usepackage{verbatim}
\usepackage{graphicx}

\usepackage{xcolor}
\usepackage{hyperref}

\usepackage{multirow}
\usepackage{graphicx}
\usepackage{booktabs}
\usepackage{float}

\usepackage[normalem]{ulem}
\useunder{\uline}{\ul}{}
\usepackage{enumitem}

\usepackage{threeparttable}

\usepackage{subfigure}
\usepackage{pgfplots}
\usepackage{tikz}
\usepackage{tikz-qtree}

\usetikzlibrary{matrix}
\usetikzlibrary{patterns}
\usepgfplotslibrary{groupplots}

\definecolor{1}{RGB}{102,102,255}
\definecolor{2}{RGB}{102,206,245}
\definecolor{3}{RGB}{255,247,102}
\definecolor{4}{RGB}{255,178,102}
\definecolor{11}{RGB}{146,209,79}
\definecolor{5}{RGB}{194,209,163}
\definecolor{6}{RGB}{102,163,255}
\definecolor{7}{RGB}{163,223,196}
\definecolor{8}{RGB}{194,163,163}
\definecolor{9}{RGB}{194,102,163}
\definecolor{10}{RGB}{255,102,102}
\definecolor{red1}{RGB}{199,21,133}
\definecolor{red2}{RGB}{255,102,102}
\definecolor{red3}{RGB}{255,182,193}
\definecolor{blue1}{RGB}{25,25,112}
\definecolor{blue2}{RGB}{30,144,255}
\definecolor{blue3}{RGB}{135,206,250}

\usepackage{tcolorbox}
\newtcolorbox{outbox}[1]{colback=blue!5!white,colframe=blue!75!black,fonttitle=\bfseries,title=#1}

\newtcolorbox{mybox}[1]{colback=red!5!white,colframe=red!75!black,fonttitle=\bfseries,title=#1}

\usepackage[linesnumbered,ruled,vlined]{algorithm2e}

\AtBeginDocument{%
  }

\setcopyright{acmlicensed}
\copyrightyear{2018}
\acmYear{2018}
\acmDOI{XXXXXXX.XXXXXXX}

\acmJournal{JACM}
\acmVolume{37}
\acmNumber{4}
\acmArticle{111}
\acmMonth{8}

\begin{document}

\title{Causal-Invariant Cross-Domain Out-of-Distribution Recommendation}

\author{Jiajie Zhu}
\email{jiajie.zhu@mq.edu.au}
\orcid{0000-0001-8673-1477}
\author{Yan Wang}
\authornote{Yan Wang is the corresponding author.}
\email{yan.wang@mq.edu.au}
\orcid{0000-0002-5344-1884}
\affiliation{%
  \institution{Macquarie University}
  \city{Sydney}
  \state{NSW}
  \country{Australia}
}

\author{Feng Zhu}
\affiliation{
  \institution{Coupang Group}
  \country{USA}}
\email{fezhu2@coupang.com}
\orcid{0000-0003-4200-0423}

\author{Pengfei Ding}
\email{pengfei.ding@mq.edu.au}
\orcid{0000-0002-7048-7518}
\author{Hongyang Liu}
\email{hongyang.liu2@hdr.mq.edu.au}
\orcid{0000-0002-4201-1934}
\affiliation{%
  \institution{Macquarie University}
  \city{Sydney}
  \state{NSW}
  \country{Australia}
}

\author{Zhu Sun}
\affiliation{%
  \institution{Singapore University of Technology and Design}
  \city{Singapore}
  \country{Singapore}}
\email{zhu_sun@sutd.edu.sg}
\orcid{0000-0002-3350-7022}

\renewcommand{\shortauthors}{Zhu et al.}

\begin{abstract}
  Cross-Domain Recommendation (CDR) aims to leverage knowledge from a relatively data-richer source domain to address the data sparsity problem in a relatively data-sparser target domain. While CDR methods need to address the distribution shifts between different domains, i.e., cross-domain distribution shifts (CDDS), they typically assume independent and identical distribution (IID) between training and testing data within the target domain. However, this IID assumption rarely holds in real-world scenarios due to single-domain distribution shift (SDDS). The above two co-existing distribution shifts lead to out-of-distribution (OOD) environments that hinder effective knowledge transfer and generalization, ultimately degrading recommendation performance in CDR. To address these co-existing distribution shifts, we propose a novel \textbf{C}ausal-\textbf{I}nvariant \textbf{C}ross-\textbf{D}omain \textbf{O}ut-of-distribution \textbf{R}ecommendation framework, called CICDOR. In CICDOR, we first learn dual-level causal structures to infer domain-specific and domain-shared causal-invariant user preferences for tackling both CDDS and SDDS under OOD environments in CDR. Then, we propose an LLM-guided confounder discovery module that seamlessly integrates LLMs with a conventional causal discovery method to extract observed confounders for effective deconfounding, thereby enabling accurate causal-invariant preference inference. Extensive experiments on two real-world datasets demonstrate the superior recommendation accuracy of CICDOR over state-of-the-art methods across various OOD scenarios.
\end{abstract}



\maketitle

\section{Introduction}

\begin{figure*}[ht]
\centering
\includegraphics[scale=0.45]{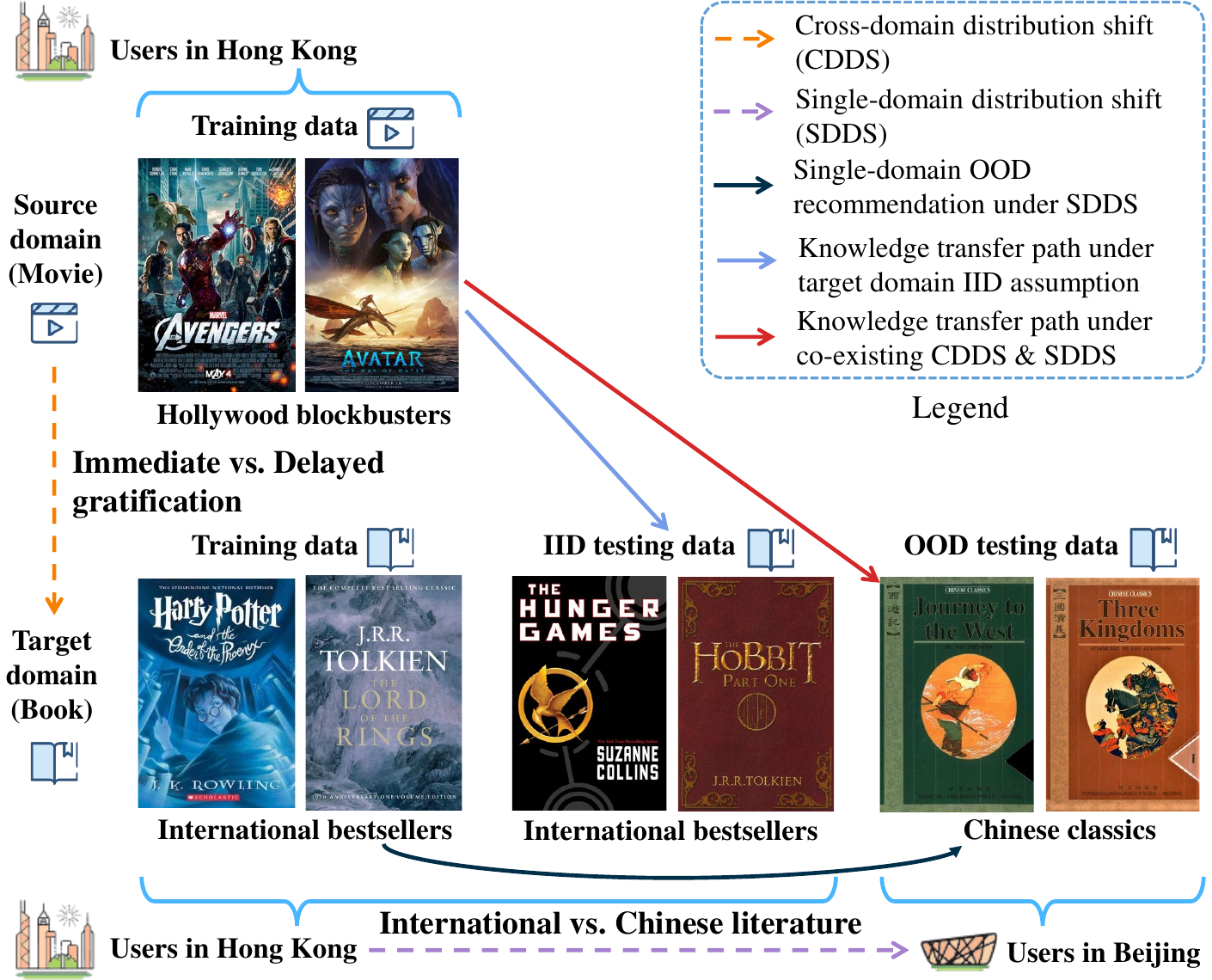} 
\vspace{-0.1in}
\caption{An illustrative example of co-existing distribution shifts (CDDS and SDDS) in CDR.}
\label{motivating_example}
\vspace{-0.2in}
\end{figure*}

To address the data sparsity problem, Cross-Domain Recommendation (CDR) leverages knowledge from a data-richer \emph{source domain} to enhance recommendation accuracy in a data-sparser \emph{target domain} \cite{zhu2018deep,ijcai2021p639}. Due to the inherent differences between domains, CDR methods need to address the distribution shifts between different domains. Meanwhile, most existing CDR methods adopt the independent and identically distributed (IID) assumption, positing that training and testing data share the same distribution in the target domain \cite{wang2022causal}. This assumption simplifies the process of problem formulation and theoretical reasoning, enabling the model to better generalize knowledge learned from training data to testing data within the target domain \cite{li2024cross}.

However, this assumption often fails to hold in real-world CDR scenarios. In CDR, recommendation performance is affected by two types of distribution shifts in terms of user preferences, interaction patterns, and item features \cite{zhang2023connecting}: \emph{cross-domain distribution shift} (CDDS) and \emph{single-domain distribution shift} (SDDS). (\textup{i}) CDDS results from data distribution differences between source and target domains \cite{li2021debiasing}. (\textup{ii}) SDDS results from data distribution differences due to temporal and regional variations in the target domain \cite{he2022causpref}, leading to mismatches between the training and testing data distributions.

Fig. \ref{motivating_example} illustrates these two types of distribution shifts in CDR. Consider an online platform that provides reviews for both movie and book domains. Specifically, in this scenario:

\noindent$\bullet$ \textbf{CDDS:} The inherent differences between domains lead to CDDS. In the movie domain, users often seek immediate gratification through visually engaging Hollywood blockbusters (e.g., \emph{Avengers} and \emph{Avatar}), while the book domain offers a slower, more immersive experience, catering to users who value delayed gratification.

\noindent$\bullet$ \textbf{SDDS:} In the book domain, regional variations in user preferences result in SDDS. Users in Hong Kong, which is a highly internationalized and modern metropolitan city, tend to reflect openness to international cultural influences and favor internationally renowned literature (e.g., international bestsellers) such as \emph{Harry Potter} and \emph{The Lord of the Rings}. In contrast, users in Beijing, known as China's cultural and political center, tend to prefer Chinese classics like \emph{Journey to the West} and \emph{Three Kingdoms}.

\noindent In such a scenario, reliable book recommendations for users in Beijing require: (1) Mitigating CDDS by transferring knowledge from movie domain to book domain, and (2) Mitigating SDDS by generalizing knowledge within book domain from users in Hong Kong to users in Beijing. While existing approaches focus on handling either CDDS or SDDS in isolation, we observe that beyond learning invariant knowledge under SDDS within book domain, there exists a knowledge transfer path that can capture invariant knowledge under both CDDS and SDDS (illustrated by the red arrow). For example, preferences for heroic adventure stories remain invariant under both CDDS and SDDS. Such invariant preferences learned from movie domain training data (\emph{The Avengers} and \emph{Avatar}) can be transferred to book domain to help make recommendation (e.g., \emph{Journey to the West}) for users in Beijing, as these works share similar narrative patterns of heroic quests and supernatural adventures. This invariant knowledge transfer path helps improve book domain recommendation performance by avoiding the potential information loss and error accumulation that could occur when sequentially applying existing approaches that handle CDDS and SDDS in isolation. The co-existence of these distribution shifts leads to complex OOD environments in CDR \cite{ding2025few}. This observation motivates us to propose a new setting of \emph{cross-domain OOD recommendation}, which simultaneously addresses two types of distribution shifts through unified modeling.

Nevertheless, this new setting presents a critical challenge that has not been fully addressed in the literature, i.e., \textbf{CH1}: \emph{How to simultaneously address cross-domain and single-domain distribution shifts to achieve reliable recommendation under OOD environments in CDR?} For CDDS, existing approaches \cite{guo2023dan,su2024cross} mainly align shared attributes, user behaviors, or auxiliary data between domains to mitigate distribution shifts \cite{zhang2025cold}. In contrast, to address SDDS, existing approaches either formulate such distribution shift as a debiasing problem, particularly targeting specific biases such as conformity bias \cite{zheng2021disentangling} or popularity bias \cite{zhao2023disentangled}, or leverage causal discovery to jointly model the causal structure and invariant user preferences \cite{he2022causpref}. However, most existing approaches are constrained to handling SDDS or CDDS in isolation, failing to handle two types of distribution shifts simultaneously, thus leading to degraded recommendation performance in CDR.

Moreover, effectively addressing the co-existing distribution shifts requires leveraging the invariance principle \cite{buhlmann2020invariance,oberst2021regularizing}, which focuses on identifying invariant relationships across distributions, particularly under OOD environments. Recent studies \cite{he2022causpref,li2024cross} have proven that such invariant relationships can be effectively represented through a causal structure between user attributes and preferences, as this structure remains invariant across distribution shifts. However, in the context of recommender systems (RSs), confounding factors, i.e., confounders in causal inference \cite{gao2022causal,zhu2023causal}, that simultaneously influence both the treatment (user preferences) and outcome (user-item interactions) \cite{zhu2025causal}, introduce confounding biases into the learned causal structure, thereby undermining the accuracy of causal-invariant user preferences inferred from such causal structure.

Furthermore, there is another challenge \textbf{CH2}: \emph{How to extract observed confounders from user reviews to further enhance the accuracy of causal-invariant user preference learning?} Existing OOD recommendation approaches either assume the absence of confounders \cite{he2022causpref,li2024cross}, or only address unobserved confounders by implicitly modeling them as external environments \cite{wang2022invariant}, overlooking the necessity of explicitly extracting and then deconfounding observed confounders. Conventional causal discovery methods provide theoretically-grounded approaches to identifying causal relationships between variables, but they require pre-defined variables and cannot automatically extract causal variables from unstructured text \cite{khatibi2024alcm}. In contrast, recent studies have demonstrated promising results in utilizing large language models (LLMs) to extract causal variables from unstructured text \cite{chen2024unlocking,causalcoat2024}. While these successes suggest the potential of extracting observed confounders from user reviews, LLMs may extract inaccurate confounders without proper theoretical guidance \cite{abdali2023extracting}. Thus, effectively extracting accurate observed confounders from user reviews to enhance causal-invariant preference learning has not been fully explored.

To address the above two challenges, we propose a novel \textbf{C}ausal-\textbf{I}nvariant \textbf{C}ross-\textbf{D}omain \textbf{O}ut-of-distribution \textbf{R}ecommendation framework, called CICDOR. To address \textbf{CH1}, we propose a dual-level causal preference learning module. This module first leverages a user preference disentanglement module to extract domain-specific and domain-shared user preferences. In addition, dual-level causal structures, represented as Directed Acyclic Graphs (DAGs), are learned for both domain-specific and domain-shared levels of modeling. Based on these causal structures, the corresponding causal-invariant user preferences are inferred at each level, thus tackling CDDS and SDDS simultaneously under OOD environments in CDR. To address \textbf{CH2}, we propose an LLM-guided confounder discovery module. This module first employs an LLM to extract candidate interaction-related causal variables from user reviews and transform them into structured data. The structured data are then fed into a conventional causal discovery method (i.e., the FCI algorithm) to uncover their underlying causal relationships, with conditional independence tests used to eliminate redundant variables. The observed confounders are first identified by leveraging the LLM and then stored in the confounder pool. Next, when such remaining variables cannot fully explain the user-item interactions, feedback is constructed through the LLM to guide the discovery of additional causal variables, forming an iterative process that continuously refines the confounder discovery. Finally, a confounder selection function is used to control confounders from the confounder pool, enabling effective deconfounding via backdoor adjustment, thereby ensuring accurate causal-invariant user preference inference.

The primary contributions of this work are as follows:
\begin{itemize}
    \item We propose CICDOR, a novel framework that discovers invariant causal structures across distributions to tackle both CDDS and SDDS for reliable cross-domain OOD recommendation;
    \item We propose a dual-level causal preference learning module that can effectively infer domain-specific and domain-shared causal-invariant user preferences, respectively;
    \item We propose an LLM-guided confounder discovery module that seamlessly integrates LLMs with a conventional causal discovery method to extract observed confounders for effective deconfounding;
    \item We conduct extensive experiments on two real-world datasets. The experimental results demonstrate that CICDOR outperforms the best-performing state-of-the-art baseline model across various OOD settings with an average increase of 6.34\% and 9.75\% w.r.t. HR@10 and NDCG@10, respectively.
\end{itemize}

\section{Related Work}
\subsection{Cross-Domain Recommendation}
CDR aims to improve recommendation performance in a relatively data-sparser target domain by employing knowledge from a relatively data-richer source domain. The existing CDR approaches can be broadly categorized into four types, namely, (1) conventional approaches, (2) causal-based approaches, (3) disentanglement-based approaches, and (4) alignment-based approaches. \emph{Conventional approaches} mainly focus on leveraging explicit content features (e.g., user/item attributes, textual data) \cite{kanagawa2019cross}, mapping user/item embeddings \cite{hu2019transfer,fu2019deeply} or rating patterns \cite{ijcai2019p587}, and integrating representations through transfer layers or by sharing network parameters \cite{hu2018conet,liu2020cross}. Furthermore, \emph{causal-based approaches} first construct causal graphs to identify confounders \cite{gao2022causal,zhu2023causal} that influence both user preferences and user-item interactions \cite{zhu2025causal}. Then, techniques like inverse propensity weighting (IPW) \cite{sato2020unbiased} or causal intervention \cite{wang2021deconfounded} are applied to debias or deconfound such confounders, ensuring that the transferred knowledge accurately reflects true user preferences and strengthens recommendation robustness. By contrast, \emph{disentanglement-based approaches} utilize techniques such as variational autoencoders (VAEs) \cite{cao2022disencdr,zhu2023domain} or introduce supervision signals such as adversarial learning \cite{wang2019recsys,su2022cross} and contrastive learning \cite{cao2023towards} to decouple domain-invariant (shared) user preferences, which remain stable across domains and capture fundamental behavioral patterns. In addition, \emph{alignment-based approaches} mitigate distribution shifts by directly aligning user behaviors and shared user-item interaction patterns \cite{zhao2023cross,xie2024heterogeneous}, or mapping auxiliary features into shared latent spaces \cite{liu2021leveraging,xie2022contrastive}, thus facilitating effective knowledge transfer. However, these approaches either overlook distribution shifts entirely, or solely focus on addressing CDDS while neglecting SDDS, limiting their effectiveness in OOD scenarios.

\subsection{Out-of-Distribution Recommendation}
OOD generalization has emerged as a fundamental research direction in artificial intelligence, with significant developments across various fields such as natural language processing \cite{borkan2019nuanced}, computer vision \cite{blanchard2021domain}, and others. Among these fields, RSs face particularly challenging OOD scenarios due to the inherently dynamic nature of user behaviors and item distributions. Existing OOD recommendation methods can be broadly categorized into three classes, i.e., (1) disentanglement-based methods, (2) causal-based methods, and (3) invariant learning-based methods. \emph{Disentanglement-based methods} aim to learn disentangled representations that separate stable user preferences from dynamic factors. For example, DICE \cite{zheng2021disentangling} disentangles user behaviors into interest and conformity components, training separate embeddings with cause-specific data to capture each factor independently. Moreover, DCCL \cite{zhao2023disentangled} employs contrastive learning to disentangle user interests from conformity behaviors, integrating item popularity signals to address data sparsity and enhance the robustness of causal representations. In contrast, \emph{causal-based methods} focus on modeling the invariant causal mechanisms in recommendation scenarios. For instance, COR \cite{wang2022causal} views user attribute changes as interventions and formulates OOD recommendation as a post-intervention inference problem, using a variational framework to model the causal relationships between user features and interactions. In addition, CausPref \cite{he2022causpref} learns causal structures through DAGs, combining invariant user preference learning with anti-preference sampling to handle implicit feedback. Different from the above two classes, \emph{invariant learning-based methods} aim to identify stable patterns across different environments while eliminating spurious correlations that may vary across distributions. For example, InvPref \cite{wang2022invariant} identifies heterogeneous environments from interaction data to separate invariant user preferences from variant ones that are affected by environmental factors, i.e., unobserved confounders. Furthermore, InvCF \cite{zhang2023invariant} learns preference representations that remain stable under popularity shifts, utilizing an auxiliary classifier to separate invariant user preferences from dynamic popularity factors. However, these existing methods either treat OOD recommendation as a debiasing problem by considering specific factors (e.g., conformity and popularity) as individual confounders, or regard environmental factors as unobserved confounders while overlooking the impact of observed confounders. This incomplete consideration of confounders limits their ability to learn robust representations for OOD recommendation.

\subsection{Causal Structure Learning}
Causal structure learning seeks to uncover causal relationships among variables from observational data, playing a fundamental role in fields such as healthcare, economics, and RSs \cite{he2022causpref}. To achieve this goal, causal discovery provides systematic approaches to inferring such relationships through statistical analysis \cite{khatibi2024alcm}. Conventional causal discovery methods, which are primarily data-driven, can be broadly categorized into dependency-based, function-based, and optimization-based methods. Dependency-based methods rely on statistical strategies to infer causal structures, but are often limited by the Markov equivalence class, making it challenging to determine unique causal directions \cite{Hasan2023ASO}. In contrast, function-based methods such as LiNGAM \cite{shimizu2006linear} utilize asymmetries in the data generation process to identify unique causal structures. In addition, optimization-based methods like NOTEARS \cite{zheng2018dags} reformulate structure learning as a continuous optimization problem, using gradient-based techniques to learn causal relationships.

\begin{table}[ht]
\setlength{\abovecaptionskip}{0cm}
\setlength{\belowcaptionskip}{0cm}
\caption{Important notations.}
\label{tab:important_notation}
\centering
\begin{tabular}{cc}
\toprule[0.7pt]
\textbf{Symbol}                      & \textbf{Definition}                    \\ \midrule[0.5pt]
$\mathcal{U}$         & the set of users  \\ 
$\mathcal{V}$         & the set of items          \\
$m$         & the number of users          \\
$n$         & the number of items          \\
$R$         & the user reviews  \\
${\mathbf{Y}}$         & the interaction matrix  \\
$y_{ij}$         & the interaction of user ${u}_i$ on item ${v}_j$          \\
$\hat y_{ij}$         & the predicted user-item interaction          \\
$k$              & the embedding dimension      \\
${*^s}$, ${*^t}$  & the notation for source domain and target domain, respectively\\
$\mathbf{E}^{inv}$   & the causal-invariant user preferences\\ 
$\mathbf{E}_{sha}$          & the domain-shared user preferences \\
$\mathbf{E}_{spe}$   & the domain-specific user preferences \\ 
$\mathbf{E}_{att}$   & the user attributes\\ 
$\mathbf{E}_u^{*}$  & the comprehensive user preferences \\
$\mathbf{E}_v$  & the item embeddings \\
$\mathbf{A}$   & the weighted adjacency matrix, i.e., DAG\\ 
$\Phi$ & the LLM \\
$\rho^\tau$ & the prompt used at the $\tau$-th round \\
$\mathcal{F}$ & the causal discovery algorithm \\
$\xi$ & the causal feedback operation \\
$\hat{R}^\tau$ & the subset of review samples used at the $\tau$-th round \\
$\mathcal{Z}^{\leq\tau}$ & the set of variables proposed at the first $\tau$-th round \\
$\mathcal{Z}_{f}^{\leq \tau}$ & the filtered variable set at the first $\tau$-th round \\
$\mathcal{Z}_{MB}^{\leq \tau}$ & the refined variable set at the first $\tau$-th round \\
$\mathcal{Z}_{pool}$ & the variable pool \\
$MB(Y)$ & the Markov Blanket of the target variable $Y$ \\
$\mathcal{C}_{pool}$ & the confounder pool \\
$\mathcal{C}$   & the confounder subspace\\
$\mathcal{G}^{\tau}$ & the causal structure discovered at the $\tau$-th round over $\mathcal{Z}_{f}^{\leq \tau} \cup \{Y\}$\\
$q \in \{-1, 0, 1\}$ & the annotation result \\
$\mathcal{Q}^{\leq\tau}$ & the data matrix of annotated values of variables proposed at the first $\tau$-th round \\
$J$         & the number of cluster centroids\\ 
$p(c)$     & the uniform distribution for prior probability\\
$\alpha $ & the weight of causal structure learning loss\\
$\beta$  & the weight of different terms of total loss\\
$\mathbf{W}$  & the weight matrix\\
\bottomrule[0.7pt]
\end{tabular}
\vspace{-0.15in}
\end{table}

Furthermore, recent advances in LLMs have led to a new paradigm for causal discovery through LLM-guided methods \cite{le2024multi,takayama2024integrating}. In contrast to conventional causal discovery methods, these methods employ the semantic understanding of variables to infer causality. Existing studies in this direction primarily follow two streams: one that explores direct causal inference through LLM prompting \cite{kiciman2023causal,zevcevic2023causal}, and another that integrates LLMs into conventional causal discovery methods as prior or posterior knowledge to improve their effectiveness \cite{ban2023causal,vashishtha2023causal}. However, existing methods either require pre-defined variables or domain expertise, or simply utilize LLMs' world knowledge as a substitute for expert reasoning. Few attempts \cite{causalcoat2024,yang2026multistage} have been made to seamlessly combine the strengths of both conventional and LLM-guided causal discovery methods for extracting observed confounders from user reviews.

\begin{figure}[ht]
\centering
\includegraphics[scale=0.42]{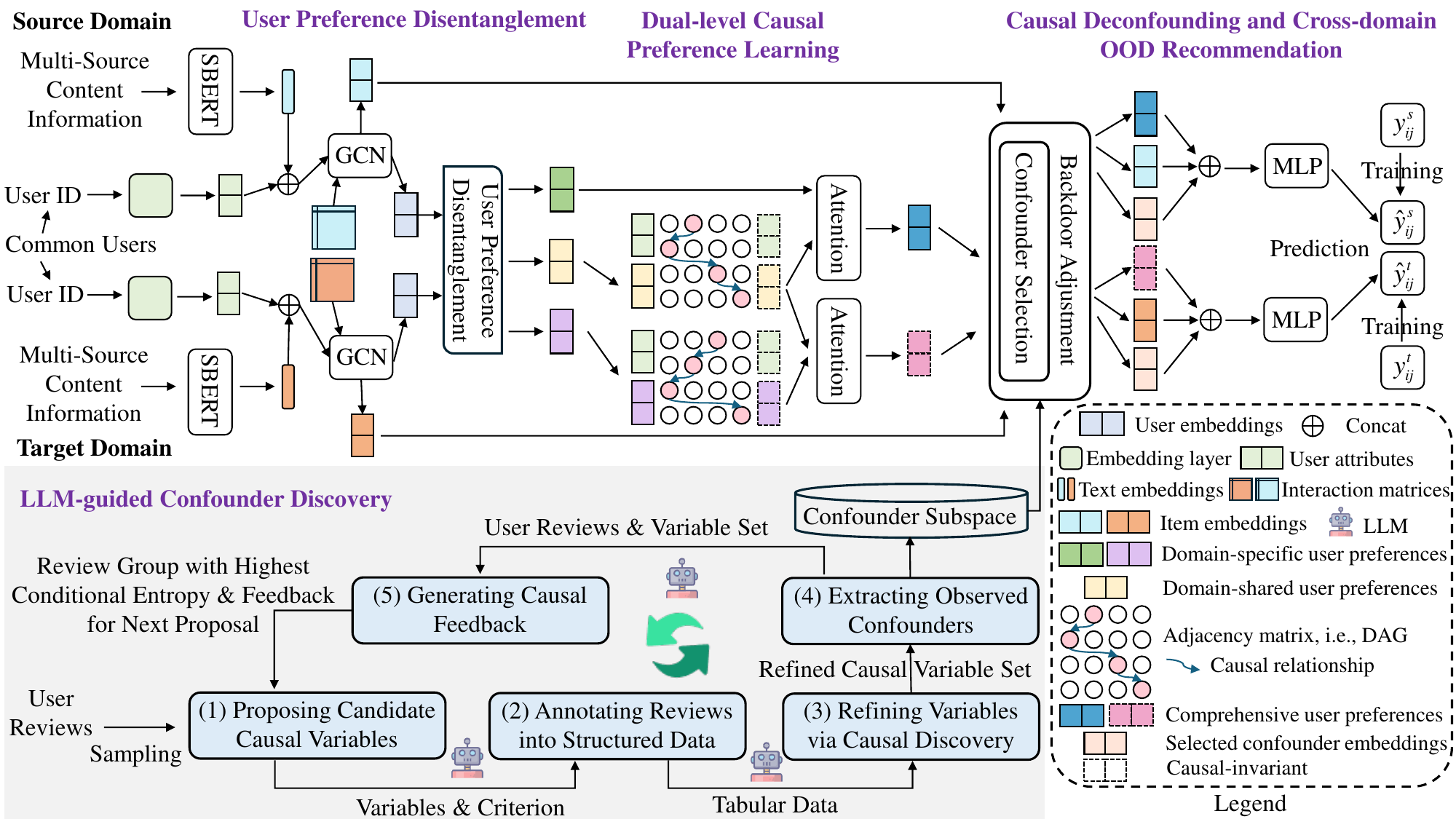} 
\vspace{-0.15in}
\caption{The flowchart of the proposed framework.}
\label{flowchart}
\vspace{-0.2in}
\end{figure}

\section{Preliminaries}

\subsection{Notations and Background}
Table \ref{tab:important_notation} summarizes the key notations employed throughout this paper for better clarity. This paper considers a source domain ${D^s}$ with relatively richer data and a target domain ${D^t}$ with sparser data. Both domains share a overlapping user set $\mathcal{U}$ of size $m = |\mathcal{U}|$. We denote the user reviews in $D^s$ and $D^t$ as $R^s$ and $R^t$, respectively. The item sets in the source and target domains are $\mathcal{V}^s$ (of size $n^s = |\mathcal{V}^s|$) and $\mathcal{V}^t$ (of size $n^t = |\mathcal{V}^t|$) respectively, with corresponding user-item interaction matrices $\mathbf{Y}^s$ and $\mathbf{Y}^t$. Here, each entry $y_{ij} \in {\{0,1\}}$ represents whether user $i$ has interacted with item $j$.

\subsection{Problem Formulation}
In this paper, we propose a new setting of cross-domain OOD recommendation.

\noindent$\bullet$ \textbf{Input:} Data from source domain $D^s$ and training data from target domain $D^t$, with data distributions $\mathbb{P}^s(\mathcal{U}, \mathcal{V}^s) \neq \mathbb{P}_{tr}^t(\mathcal{U}, \mathcal{V}^t)$.

\noindent$\bullet$ \textbf{Output:} A cross-domain recommender system that can accurately predict whether any user $i \in \mathcal{U}$ interacts with an item $j \in \mathcal{V}^t$ (i.e., $\hat{y}_{ij}^t$) in the target domain testing data $D_{te}^t$, where $\mathbb{P}_{tr}^t(\mathcal{U}, \mathcal{V}^t) \neq \mathbb{P}_{te}^t(\mathcal{U}, \mathcal{V}^t)$.

\subsection{Connection with Existing Works and Our Novel Insights}
Our proposed CICDOR is the first work to explicitly propose the new setting of \emph{cross-domain OOD recommendation} and tackle both cross-domain distribution shifts (CDDS) and single-domain distribution shifts (SDDS) through unified causal modeling. CICDOR not only builds on existing works, but also contributes several novel insights \cite{ding2025few,zhu2025causal}. Below, we analyze key gaps in existing literature and highlight how CICDOR addresses these gaps through novel insights:

\begin{itemize}[leftmargin=*]
\item \textbf{Treating CDDS and SDDS as independent problems:} Most existing approaches are constrained to handling CDDS and SDDS in isolation, failing to recognize that their co-existence gives rise to complex OOD environments in CDR. In contrast, CICDOR addresses this gap by adopting a unified causal modeling perspective that jointly accounts for both types of distribution shifts.

\item \textbf{Overlooking observed confounder extraction from unstructured text:}
Existing OOD recommendation methods either assume no confounders or focus solely on unobserved ones, overlooking the extraction of observed confounders from user reviews. Meanwhile, LLM-based causal variable extraction frameworks such as COAT \cite{causalcoat2024} demonstrate the feasibility of discovering causal variables from unstructured text, but are designed for general-purpose causal analysis rather than recommendation-specific confounder extraction. CICDOR adapts these validated discovery components with domain-specific reasoning to extract both single-domain and cross-domain confounders from user reviews, enabling effective deconfounding for reliable cross-domain OOD recommendation.

\item \textbf{Relying on structural criteria in confounder extraction without semantic understanding:} While conventional confounder detection methods \cite{vanderweele2019principles, williamson2014introduction} can theoretically identify confounders from causal structures, they operate purely on structural criteria without semantic context, returning only confounder names without descriptions. In contrast, CICDOR combines LLM-guided extraction with conventional causal discovery methods to obtain a theoretically grounded, refined set of candidate variables, from which observed confounders are further extracted with detailed descriptions and reasoning processes. Although backdoor adjustment itself is not our contribution, our novelty lies in seamlessly combining the strengths of conventional and LLM-guided approaches to identify semantically rich observed confounders from user reviews, enabling effective deconfounding for learning debiased causal-invariant user preferences.
\end{itemize}

\section{Methodology}
\subsection{Framework Overview}
In this section, we propose a novel \textbf{C}ausal-\textbf{I}nvariant \textbf{C}ross-\textbf{D}omain \textbf{O}ut-of-distribution \textbf{R}ecommendation framework, called CICDOR. As illustrated in Fig. \ref{flowchart}, CICDOR is built upon dual-level causal structures and follows the principle of causal invariance to tackle the OOD problem in CDR. Specifically, (1) To simultaneously address CDDS and SDDS, CICDOR proposes a dual-level causal preference learning module that discovers causal structures and infers causal-invariant user preferences at both domain-specific and domain-shared levels. (2) To ensure accurate causal-invariant user preference inference, CICDOR proposes an LLM-guided confounder discovery module. This module identifies interaction-related causal variables from reviews and further extracts observed confounders for effective deconfounding, thus enabling the accurate inference of debiased comprehensive causal-invariant user preferences for reliable cross-domain OOD recommendation.

\subsection{User Preference Disentanglement} 
\subsubsection{\textbf{Embedding Construction}}
Taking the target domain $D^t$ as an example, let $\mathbf{u} \in \{0,1\}^m$ and $\mathbf{v}^t \in \{0,1\}^{n^t}$ denote the one-hot vectors of users and items, respectively \cite{zhao2022multi}. We first transform user one-hot vectors into $k$-dimensional attribute embeddings\footnote{Unless otherwise specified, all embedding dimension mentioned hereafter is k.} via $\mathbf{E}_{att}^t = \mathbf{W}_{att}^t\mathbf{u}$, where $\mathbf{W}_{att}^t \in \mathbb{R}^{k \times m}$ is a learnable embedding matrix. To enhance the extraction of disentangled user preferences, we leverage multi-source content information including user reviews and item details \cite{zhu2020graphical}. Specifically, we collect all reviews written by user $u_i$ into a user document, while for item $v_j$, we collect both its details and associated reviews into an item document \cite{zhu2025causal}. Next, we employ a sentence transformer model to encode these documents, converting them into dense representations. This process generates user text embeddings $\mathbf{E}_{ut}^t$ and item text embeddings $\mathbf{E}_{vt}^t$ for all users and items in the training set. We then concatenate $\mathbf{E}_{att}^t$ and $\mathbf{E}_{ut}^t$ to obtrain the combined user embeddings $\mathbf{E}_{uc}^t$. These text embeddings are then projected into fixed-dimensional spaces through multi-layer perceptron (MLP) layers to create initial user embeddings $\mathbf{E}_{ui}^t$ and initial item embedding $\mathbf{E}_{vi}^t$ in the target domain. Likewise, we can obtain $\mathbf{E}_{att}^s$, $\mathbf{E}_{ui}^s$ and $\mathbf{E}_{vi}^s$ in the source domain as well. Using these initial embeddings and the corresponding interaction matrices as inputs, we construct two heterogeneous graphs and then employ the Graph Convolutional Network (GCN) \cite{KipfW17} to generate enriched user embeddings $\mathbf{E}_u$ and item embeddings $\mathbf{E}_v$ that capture both content semantics and collaborative patterns within each domain.

\subsubsection{\textbf{User Preference Disentanglement}}
In CDR, a crucial challenge is to determine what knowledge should be transferred across domains \cite{ijcai2021p639,zang2022survey}. To address this challenge, we disentangle the enriched user embeddings $E_u$ to obtain transferable domain-shared user preferences while preserving domain-specific user preferences that capture unique user behaviors in each domain. Inspired by \cite{choi2022based}, we implement a user preference disentanglement module with three encoders: one domain-shared encoders and two domain-specific encoders (all using the two-layer MLP). These encoders collaboratively disentangle user preferences into domain-shared and domain-specific components.

For effective disentanglement, we employ a domain discriminator implemented as a fully-connected neural network. This discriminator attempts to identify whether encoded user preferences originate from the source or target domain by minimizing a domain classification loss. Between the domain-shared encoder and discriminator, we utilize a gradient reversal layer (GRL) \cite{ganin2015unsupervised}, which multiplies gradients by a negative constant during backpropagation. This causes the domain-shared encoder to maximize the domain classification loss, thereby learning to generate user preferences that confuse the discriminator. Through this domain adversarial training \cite{ganin2016domain}, the domain-shared encoder effectively extracts transferable preferences of overlapping users. To optimize this disentanglement, we define the domain classification losses using binary cross-entropy as follows:
\begin{equation}
\mathcal{L}_{\text{sha}}^s = -\frac{1}{N^s} \sum_{i=1}^{N^s} \log(1 - \hat{d}_{\text{sha}}^s), \quad
\mathcal{L}_{\text{sha}}^t = -\frac{1}{N^t} \sum_{i=1}^{N^t} \log(\hat{d}_{\text{sha}}^t),
\end{equation}
where $d$ denotes the binary domain label (0 for source domain and 1 for target domain) and $\hat{d}$ is the discriminator's predicted probability of the input user preferences belonging to the target domain. Similarly, the domain-specific encoders connect directly to the discriminator without GRL and are trained to preserve domain-specific features by minimizing the domain classification loss $\mathcal{L}_{\text{spe}}^s$ and $\mathcal{L}_{\text{spe}}^t$.

Then, the above losses are combined to form the overall domain disentanglement objective:
\begin{equation}
\mathcal{L}^{\text{dom}} = \gamma(\mathcal{L}_{\text{sha}}^s + \mathcal{L}_{\text{spe}}^s) + (1-\gamma)(\mathcal{L}_{\text{sha}}^t + \mathcal{L}_{\text{spe}}^t),
\end{equation}
where $\gamma$ is a balancing parameter between the source and target domain losses during disentanglement. The details of user preference disentanglement process can be found in \cite{choi2022based}. Finally, we can obtain domain-shared user preferences $\mathbf{E}_{sha}$ and domain-specific user preferences $\mathbf{E}_{spe}^s$ and $\mathbf{E}_{spe}^t$.

\subsection{Dual-level Causal Preference Learning}
After obtaining domain-specific and domain-shared user preferences, inspired by CDCOR \cite{li2024cross}, we model the causal relationships between user attributes and user preferences, aiming to identify causal structures that remain invariant across distribution shifts. While CDCOR primarily focuses on learning causal structures at the domain-shared level to assist the learning of causal relationships in the target domain, it overlooks a crucial aspect: the invariant causal structure within the target domain itself, which is fundamental for robust single-domain OOD recommendation. Learning the domain-specific causal structure effectively addresses SDDS by capturing invariant causal relationships within the target domain, while the domain-shared causal structure provides supplementary support for mitigating both SDDS and CDDS. Motivated by the complementary nature of these two causal structures, we propose a dual-level causal preference learning module that learns two DAGs to model causal relationships at each level, respectively.

As shown in Fig. \ref{flowchart}, we do not learn the source domain's domain-specific causal structure. This design decision is based on two key considerations: (1) our primary goal is to enhance OOD recommendation performance in the target domain, and the source domain is only used to provide auxiliary training signals without requiring OOD generalization; (2) source domain's domain-specific user preferences are only used within its own prediction branch and do not participate in target domain inference, making causal invariance unnecessary for these representations. This design reduces model complexity while maintaining essential causal structure learning components.

Taking the domain-specific level as an example, we represent its DAG as a weighted adjacency matrix $\mathbf{A}_{\text{spe}} \in \mathbb{R}^{2k \times 2k}$, where $A_{i,j}$ indicates the strength of the causal influence from node $i$ to node $j$. Each node represents one dimension of either user attributes or domain-specific user preferences embedding. The structural causal model (SCM) can be formulated as follows: 
\begin{equation}
\label{eq2}
    \mathbf{B}_{\text{spe}} = \mathbf{A}_{\text{spe}}^\top \mathbf{B}_{\text{spe}} + \boldsymbol{\epsilon},
\end{equation}
where $\boldsymbol{\epsilon}$ represents the noise term, and $\mathbf{B}_{\text{spe}} = \mathbf{E}_{\text{att}}^t \Vert \mathbf{E}_{\text{spe}}^t \in \mathbb{R}^{2k}$, and $\Vert$ denotes the operation of concatenation. Eq.(\ref{eq2}) defines how child nodes are determined by their parent nodes through $\mathbf{A}_{\text{spe}}$. When Eq.(\ref{eq2}) holds, $\mathbf{A}_{\text{spe}}$ represents the causal structure at the domain-specific level, capturing the invariant relationships between user attributes and domain-specific user preferences. To learn $\mathbf{A}_{\text{spe}}$, we minimize the following reconstruction loss:
\begin{equation}
\mathcal{L}_{\text{spe}}^{\text{rec}} = \frac{1}{N} \sum_{i=1}^N \|\mathbf{B}_i - \mathbf{A}_{\text{spe}}^\top \mathbf{B}_i\|_2^2.
\end{equation}
To ensure that $\mathbf{A}_{\text{spe}}$ remains acyclic, as required by DAG properties, we optimize the following loss as a constraint \cite{zheng2018dags}:
\begin{equation}
\mathcal{L}_{\text{spe}}^{\text{dag}}=\operatorname{Tr}(\mathbf{e}^{\mathbf{A} \circ \mathbf{A}}) - k,
\end{equation}
where $\operatorname{Tr}(\cdot)$ computes the trace, $\circ$ is the element-wise product, and $\mathbf{e}$ represents the Euler’s number. Moreover, to ensure $\mathbf{A}_{\text{spe}}$ aligns with the nature of preference formation, we introduce two structural constraints: (1) causal influences should only flow from user attribute nodes to user preference nodes, and (2) user preference nodes cannot be root nodes \cite{he2022causpref}. These constraints align with both causal intuition and the characteristics of preference formation. The first constraint reflects that user preferences are shaped by their inherent attributes, while the second constraint ensures that preferences are always derived from some underlying personal attributes rather than emerging spontaneously. To enforce these constraints, we optimize the following two losses:
\begin{align}
\mathcal{L}_{\text{spe}}^{\text{path}} &= \|\mathbf{A}_{[k+1:2k, 1:k]}\|_1, \\
\mathcal{L}_{\text{spe}}^{\text{root}} &= \sum_{i=k+1}^{2k} -\log \|\mathbf{A}_{[:,i]}\|_1.
\end{align}
Then, we can obtain the domain-specific causal loss, which is expressed as follows:
\begin{equation}
    \mathcal{L}_{\text{spe}}^{\text{cau}} = \mathcal{L}_{\text{spe}}^{\text{rec}} + \alpha_1 \mathcal{L}_{\text{spe}}^{\text{dag}} + \alpha_2 \mathcal{L}_{\text{spe}}^{\text{path}} + \alpha_3 \mathcal{L}_{\text{spe}}^{\text{root}} + \alpha_4 \|\mathbf{A}_{\text{spe}}\|_1.
\end{equation}
Similarly, we can obtain the domain-shared causal loss $\mathcal{L}_{\text{sha}}^{\text{cau}}$. Thus, we formulate the dual-level causal loss as follows:
\begin{equation}
\label{eq8}
\mathcal{L}^{\text{cau}} = \mathcal{L}_{\text{spe}}^{\text{cau}} + \mathcal{L}_{\text{sha}}^{\text{cau}}.
\end{equation}

When training is completed, we can infer domain-specific causal-invariant user preferences using the learned causal structure $\mathbf{A}_{\text{spe}}$. During inference, we feed $\mathbf{B}_{\text{spe}} = \mathbf{E}_{att}^t \Vert \mathbf{0}$ into Eq.(\ref{eq2}), where $\mathbf{0}$ is a zero vector replacing domain-specific user preferences. From the output $\mathbf{\hat{H}}_{\text{spe}}$, we extract its posterior $k$-dimensional vector as the domain-specific causal-invariant user preferences $\mathbf{E}_{spe}^{tinv}$ in target domain. Since $\mathbf{A}_{\text{spe}}$ captures the invariant causal mechanism of how users' attributes generate their preferences, the inferred preferences through $\mathbf{A}_{\text{spe}}$ are inherently causal-invariant and thus more reliable for OOD recommendation. Likewise, we can obtain $\mathbf{E}_{sha}^{sinv}$ and $\mathbf{E}_{sha}^{tinv}$. We then apply an attention mechanism to fuse $\mathbf{E}_{spe}^{tinv}$ and $\mathbf{E}_{sha}^{tinv}$ into comprehensive user preferences $\mathbf{E}_u^{*tinv}$, and similarly fuse $\mathbf{E}_{spe}^{s}$ and $\mathbf{E}_{sha}^{sinv}$ into $\mathbf{E}_u^{*s}$ \cite{zhu2023domain}.

\subsection{LLM-guided Confounder Discovery}
To further enhance the reliability of the inferred causal-invariant preferences, we propose to address the influence of observed confounders during the training phase. User reviews, which contain rich descriptions of user interactions, provide valuable information about various interaction-related causal variables. LLMs' strong capabilities in understanding natural language text make them particularly suitable for extracting causal variables from unstructured reviews \cite{abdali2023extracting,chen2024unlocking}. However, directly extracting confounders from reviews could be imprecise due to LLMs' potential hallucinations and the complex nature of causal relationships. To address this challenge, we leverage COAT \cite{causalcoat2024}, which provides a suitable framework for discovering and refining candidate causal variables from unstructured text by integrating LLMs with a theoretically grounded causal discovery approach. We adopt COAT because, unlike conventional causal discovery methods \cite{spirtes2001causation} that rely on predefined structured variables, it can discover semantically meaningful candidate causal variables from unstructured reviews, while its causally grounded variable refinement process offers a more reliable basis for subsequent confounder extraction compared with direct LLM-based approaches. Based on this foundation, we further extend COAT to extract observed confounders from user reviews for cross-domain OOD recommendation.

To adapt this framework to our recommendation scenario, we introduce several key adaptations for observed confounder discovery in cross-domain OOD recommendation. While COAT extracts general causal variables for interpretability analysis, our goal is to identify observed confounders that simultaneously influence user preferences and user-item interactions. First, we inject domain information into prompts to guide domain-specific reasoning, enabling our approach to naturally capture both single-domain confounders, which are only causally valid in one domain (e.g., \emph{Setup service} in the Electronics domain), and cross-domain confounders that, despite being common to both domains, remain causally active within each (e.g., \emph{Price promotion} in both Electronics and Clothing domains). Second, we develop an LLM-based confounder extraction approach that operates on a theoretically validated variable set obtained through the conventional causal discovery method, thereby reducing hallucination risks. Finally, while confounder discovery is performed independently to preserve domain-specific semantic fidelity, the extracted confounders are aligned across dimensions and clustered into subspaces, enabling an implicit cross-domain interaction mechanism: the removal of shared confounding biases jointly influences the optimization of domain-shared causal-invariant user preferences, facilitating robust cross-domain knowledge transfer.

\begin{algorithm}[H]
    \normalsize
    \caption{\textsc{LLM-guided Confounder Discovery}}
    \label{alg:llmcd}
    \LinesNumbered
    \KwIn{Target domain user reviews $R^t$, LLM $\Phi$, maximum number of rounds $\tau_{max}$, causal discovery algorithm $\mathcal{F}$, causal feedback operation $\xi$.}
    \KwOut{Target domain confounder subspace $\mathcal{C}^t$.}

    Initialize $\tau \leftarrow 1$, $\mathcal{Z}^{\leq 0} \leftarrow \emptyset$, $\mathcal{Z}_{pool} \leftarrow \emptyset$, $\mathcal{C}_{pool}^{0} \leftarrow \emptyset$\;
    
    Prepare initial prompt $\rho_{pro}^{1}$ and randomly select initial review samples $\hat{R}^{1}$\;

    \While{not converged and $\tau \leq \tau_{max}$}{
        /* Step 1: Proposing Candidate Causal Variables */\\
        $\mathcal{Z}^{\tau} \leftarrow \Phi(\rho_{pro}^{\tau}, \hat{R}^{\tau})$\;
        $\mathcal{Z}^{\leq\tau} \leftarrow \mathcal{Z}^{\leq\tau-1} \cup \mathcal{Z}^{\tau}$\;

        /* Step 2: Annotating Reviews into Structured Data */\\
        \For{each review $r_i \in R^t$}{
            \For{each variable $z_j \in \mathcal{Z}^{\tau}$}{
                $q_{i,j} \leftarrow \Phi(r_i, z_j, \rho_{ano})$\;
            }
        }
        Form data matrix $\mathcal{Q}^{\leq\tau}$ from all annotation vectors\;

        /* Step 3: Refining Variables via Causal Discovery */\\
        Filter variables from $\mathcal{Z}^{\leq\tau} \cup \mathcal{Z}_{pool}$ using CI test to obtain $\mathcal{Z}_{f}^{\leq \tau}$\;
        $\mathcal{G}^{\tau} \leftarrow \mathcal{F}(\mathcal{Z}_{f}^{\leq \tau} \cup \{Y^t\})$\;
        Extract Markov Blanket $\mathcal{Z}_{MB}^{\leq \tau} \leftarrow MB(Y^t)$ from $\mathcal{G}^{\tau}$\;
        Update $\mathcal{Z}_{pool}$ with filtered variables and variables not in $\mathcal{Z}_{MB}^{\leq \tau}$\;

        /* Step 4: Extracting Observed Confounders */\\
        \If{$\mathcal{C}_{pool}^{\tau-1} = \emptyset$}{
            Use zero-shot prompting to extract confounders\;
        }
        \Else{
            Use few-shot prompting with examples to extract confounders\;
        }
        $\mathcal{C}_{pool}^{\tau} \leftarrow \mathcal{C}_{pool}^{\tau-1} \cup \Phi(\rho_{ext}, \mathcal{Z}_{MB}^{\leq \tau})$\;

        /* Step 5: Generating Causal Feedback */\\
        $(\rho_{pro}^{\tau+1}, \hat{R}^{\tau+1}) \leftarrow \xi(\mathcal{Z}_{MB}^{\leq \tau}, R^t, \rho_{pro}^{\tau})$\;
        $\tau \leftarrow \tau + 1$\;
    }
    Encode each confounder in $\mathcal{C}_{pool}^{\tau_{max}}$ into embeddings\;
    Apply K-means clustering to obtain $J$ representative confounders to form $\mathcal{C}^t$\;

    \Return{$\mathcal{C}^t$}\;
\end{algorithm}

\begin{figure}[ht]
\centering
\includegraphics[scale=0.78]{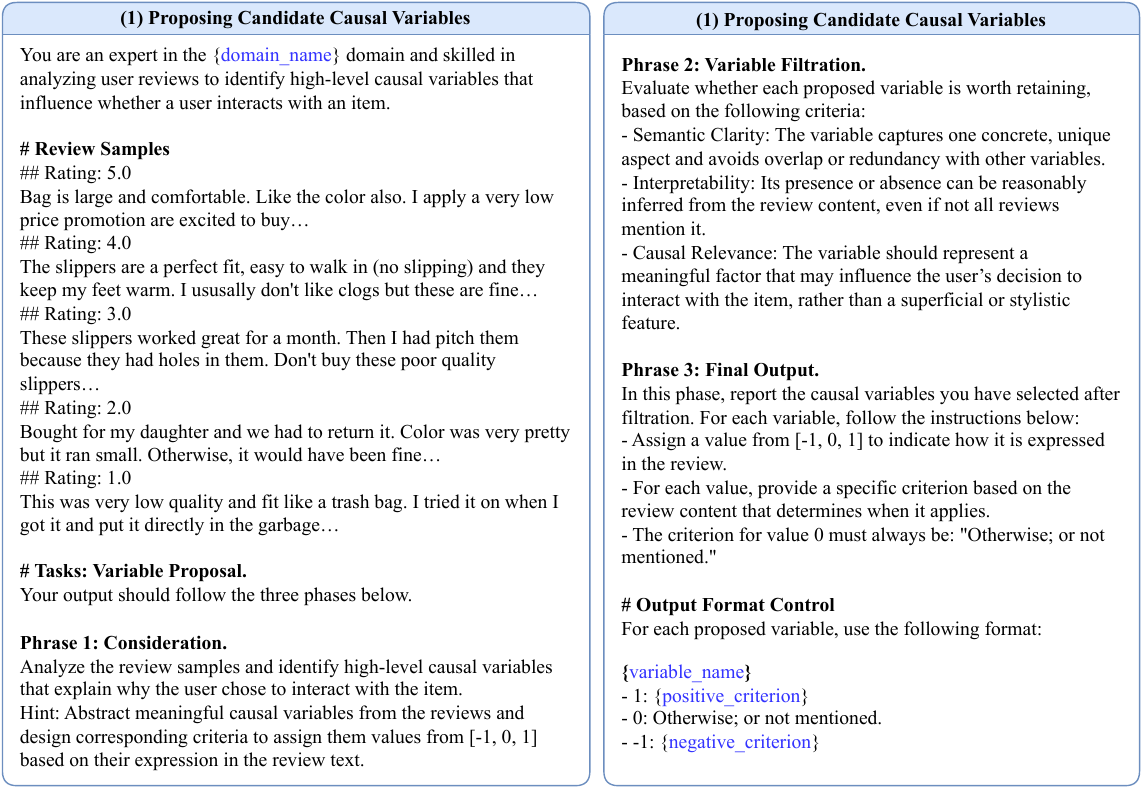} 
\vspace{-0.2in}
\caption{Illustration of the prompt template for proposing candidate causal variables.}
\label{variable_proposal}
\vspace{-0.2in}
\end{figure}

We now present the overall procedure of our proposed module. The LLM-guided confounder discovery module operates in an iterative manner. Specifically, we first use an LLM to propose candidate causal variables related to user interactions, employ a conventional causal discovery method to refine these variables, and then carefully extract observed confounders from the refined variable set. We use superscript $\tau$ to denote the $\tau$-th iteration. The module consists of five key steps. Taking the target domain $D^t$ as an example, the complete procedure is outlined in Algorithm~\ref{alg:llmcd}, and we elaborate on each step in detail in the following subsections. Likewise, the same procedure is also applied to the source domain $D^s$.

\subsubsection{\textbf{Proposing Candidate Causal Variables}}
Building upon the factor proposal of COAT framework \cite{causalcoat2024}, we adapt the prompt design to extract causal variables from user reviews. Specifically, in Fig. \ref{variable_proposal}, we design a prompt $\rho_{pro}$ for an LLM $\Phi$ to induce candidate causal variables that may affect our target variable $Y^t$ (whether a user interacts with an item in $D^t$). The prompt contains three components: review samples, task instructions, and output format control. (i) \emph{Review samples:} Target domain user reviews $R^t$ are grouped by rating scores at first. Then a few reviews are sampled from each rating group (denoted as $\hat{R} \subset R^t$) due to LLM's context length limitation. The domain name is included in the prompt alongside these samples to establish the appropriate context. These review samples, coming from users who have interacted with items ($Y^t=1$), tend to exhibit distinct patterns under different ratings, helping capture comprehensive causal variables. The domain name is provided to help the LLM $\Phi$ better interpret the context. (ii) \emph{Task instructions:} To emulate the analytical process of human experts \cite{pearl2018book}, the LLM $\Phi$ is guided to propose candidate causal variables through three phases: consideration of potential interaction-related variables, variable filtration to ensure semantic distinctness and avoid redundancy, and output of filtered variables. Each filtered variable will be presented with its name and criterion that defines positive cases, negative cases, and cases otherwise or not mentioned, corresponding to values 1, -1, and 0, respectively. (iii) \emph{Output format control:} The output format control specifies the required format for presenting the proposed variables and their criteria. The set of variables proposed in the $\tau$-th round is denoted as $\mathcal{Z}^\tau=\Phi(\rho_{pro}^\tau,\hat{R}^\tau)$. We accumulate variables across rounds to form an initial candidate variable set $\mathcal{Z}^{\leq\tau}=\mathcal{Z}^1\cup\cdots\cup \mathcal{Z}^\tau$.

\subsubsection{\textbf{Annotating Reviews into Structured Data}}
After obtaining candidate causal variables with their criteria, we apply these criteria to annotate all reviews in $R^t$ and transform them into structured data. This two-step approach (first extracting variable criteria from sample reviews, then annotating all reviews based on these criteria) offers significant advantages over directly extracting variables from the entire review set in a single step. It ensures consistent annotation standards across all reviews and allows us to progressively accumulate variables across rounds. For each review $r_i \in R^t$, we instruct LLM $\Phi$ to annotate it against each variable $z_j \in \mathcal{Z}^{\leq\tau}$ based on the variable's criterion, determining its value:
\begin{equation}
    q_{i,j} := \Phi(r_i, z_j, \rho_{ano}),
\end{equation}
where $q_{i,j} \in \{-1, 0, 1\}$ represents the annotation result, and $\rho_{ano}$ is the annotation prompt. The annotation results for each review $r_i$ form a vector $\mathbf{q}_i = [q_{i,1}, q_{i,2}, ..., q_{i,|\mathcal{Z}^{\leq\tau}|}]$. These annotation vectors collectively form a data matrix $\mathcal{Q}^{\leq\tau} := [\mathbf{q}_1^T; \mathbf{q}_2^T; ...; \mathbf{q}_{|R^t|}^T]$. As we discover new variables in additional rounds, the matrix $\mathcal{Q}^{\leq\tau}$ can be expanded with new columns, providing the structured input required for the subsequent conventional causal discovery method.

\subsubsection{\textbf{Refining Variables via Causal Discovery}}
The candidate causal variables extracted by LLMs may contain inaccuracies, as LLMs might misinterpret review content or propose variables without true causal relationships with the target variable $Y^t$. To address this challenge, we implement the variable refinement via causal discovery. Inspired by the method introduced in \cite{causalcoat2024}, our refinement begins with a filtering process using conditional independence (CI) tests. We evaluate each variable in the candidate variable set $\mathcal{Z}^{\leq\tau}$ through an iterative process. Specifically, for each candidate variable $z_j$, we test whether it maintains a significant statistical association with $Y^t$ when controlling for the set of variables already identified as relevant in previous iterations of testing. Variables that pass this test are confirmed as relevant and then retained, forming a filtered variable set $\mathcal{Z}_{f}^{\leq \tau}$, while others are eliminated. The filtering process not only removes irrelevant variables, but also helps address redundancy by eliminating variables that provide similar information as previously confirmed ones, thus providing a more reliable foundation for subsequent causal discovery.

After the initial filtering process, we feed the filtered variable in $\mathcal{Z}_{f}^{\leq \tau}$ and the target variable $Y^t$ into the Fast Causal Inference (FCI) algorithm to discover the causal structure $\mathcal{G}^{\tau}$, which can be expressed as follows.
\begin{equation}
\mathcal{G}^{\tau} = \mathcal{F}(\mathcal{Z}_{f}^{\leq \tau} \cup \{Y^t\}),
\end{equation}
where $\mathcal{F}$ denotes the FCI algorithm. FCI algorithm \cite{spirtes2001causation} is a conventional causal discovery method that accommodates potential confounders, making it well-suited for our context where causal variables proposed by LLM might be entangled with confounders \cite{jin2024can}. In addition, FCI algorithm offers flexibility regarding different functional forms of causal relationships (e.g., linear or non-linear), which corresponds well to the diverse relationship patterns we need to identify in our causal structure learning \cite{causalcoat2024}.

From the discovered causal structure $\mathcal{G}^{\tau}$, we further extract the Markov Blanket of the target variable, denoted as $MB(Y^t)$. This extraction is crucial for our variable refinement process because it allows us to focus only on the most influential variables while discarding those that provide redundant or irrelevant information, thereby achieving the optimal balance between model simplicity and effectiveness for both prediction and causal understanding of user interactions. In the causal structure, $MB(Y^t)$ consists of $Y^t$'s parents, $Y^t$'s children, and the parents of $Y^t$'s children other than $Y^t$ itself \cite{pearl2014probabilistic}. Formally, the Markov Blanket satisfies: $Y^t \perp\!\!\!\perp \mathcal{Z} | MB(Y^t)$ for any set $\mathcal{Z}$ of proposed causal variables disjoint from both $MB(Y^t)$ and $Y^t$ itself. This means that given $MB(Y^t)$, no other proposed variable provides any additional information about $Y^t$. By focusing on the Markov Blanket, our variable refinement process concludes with the identification of the refined variable set $\mathcal{Z}_{MB}^{\leq \tau}$, which is the most compact set of variables necessary and sufficient for reasoning about the target variable $Y^t$.

To avoid potentially discarding valuable variables at early stages, we maintain a variable pool $\mathcal{Z}_{pool}$ that stores both variables eliminated during the initial CI test filtering and those from $\mathcal{Z}_{f}^{\leq \tau}$ that are not included in the Markov Blanket. This mechanism ensures that valuable variables have the opportunity to be reassessed in subsequent analyses, especially when certain variables might reveal their causal relevance only under specific conditioning with other variables.

Overall, this three-step variable refinement process provides theoretical guarantees that complement the initial LLM-based variable proposal. By integrating the FCI algorithm along with supporting statistical methods, we systematically refine the proposed variables: CI test filtering removes both statistically irrelevant and redundant variables, the FCI algorithm discovers the underlying causal structure among remaining variables, and Markov Blanket extraction determines the minimal sufficient set for causal reasoning. This integration ensures that our refined variable set is theoretically grounded and contains only variables with true causal relationships to the target variable. It provides a reliable foundation for subsequent analyses, unlike approaches that directly use LLM, which may include inaccuracies.

\subsubsection{\textbf{Extracting Observed Confounders}}
After obtaining the refined variable set $\mathcal{Z}_{MB}^{\leq \tau}$ through the variable refinement process, we proceed to extract observed confounders using LLMs. In the context of RSs, confounders are defined as causal variables that simultaneously influence both the treatment (typically manifested as user preferences) and the outcome (represented by user-item interactions) \cite{zhu2025causal}. If not properly addressed, these confounders will introduce confounding biases into user preference modeling, thus reducing the performance of CDR.

Direct extraction of confounders from raw reviews using LLMs presents several limitations. Specifically, it risks identifying false confounders without theoretical validation, remains vulnerable to the hallucination problem of LLMs, and lacks guarantees of causal relevance to the target variable. To address these limitations, we propose to extract confounders from the refined variable set $\mathcal{Z}_{MB}^{\leq \tau}$, which contains only variables with theoretically validated causal relationships to the target variable. Inspired by the method introduced in \cite{abdali2023extracting}, we develop a Chain-of-Thought (CoT) prompting strategy instructing the LLM $\Phi$ to analyze each variable in $\mathcal{Z}_{MB}^{\leq \tau}$ in a step-by-step manner. This strategy is particularly valuable for confounder extraction as determining causal influence patterns requires complex judgment that benefits from explicit intermediate reasoning steps. 

\begin{figure}[ht]
\centering
\includegraphics[scale=0.78]{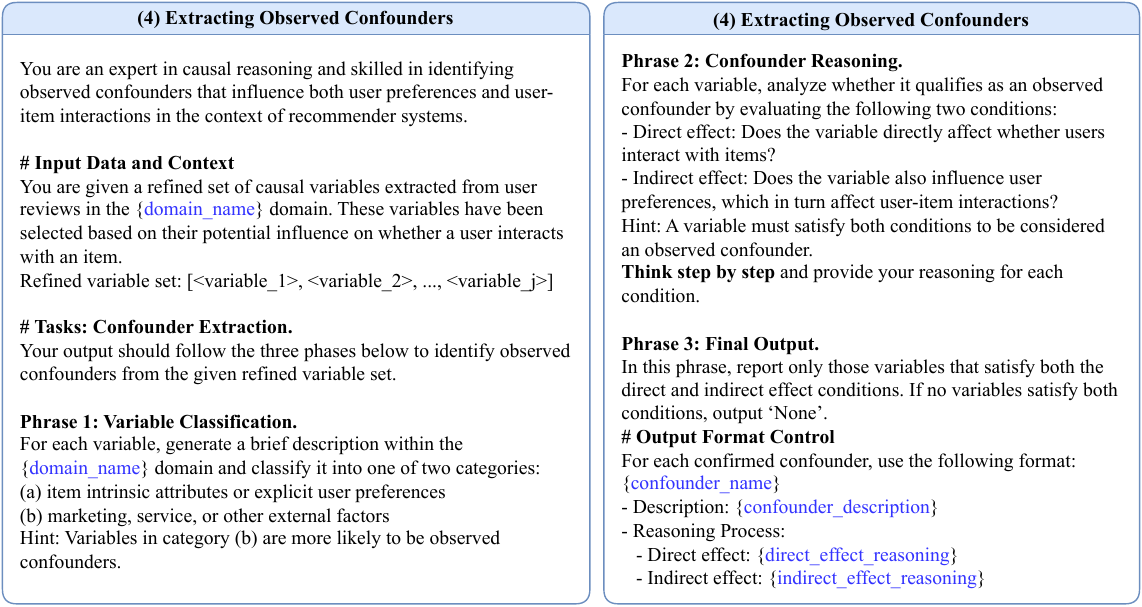} 
\vspace{-0.2in}
\caption{Illustration of the prompt template for extracting observed confounders using zero-shot prompting.}
\label{confounder_extraction}
\vspace{-0.2in}
\end{figure}

As illustrated in Fig. \ref{confounder_extraction}, our prompt $\rho_{ext}$ comprises three essential components to facilitate effective confounder extraction using an LLM $\Phi$: (i) \emph{Input data and context}, which provides the LLM $\Phi$ with the refined variable set $\mathcal{Z}_{MB}^{\leq \tau}$ and the target domain name, enabling the accurate identification of observed confounders within the given context. (ii) \emph{Task instructions}, which guide the LLM $\Phi$ to first generate a brief description for each variable in $\mathcal{Z}_{MB}^{\leq \tau}$, and then employ CoT reasoning to analyze each variable systematically. To facilitate efficient reasoning, the analysis includes a preliminary classification of variables into two categories: (a) item intrinsic attributes or explicit user preferences, and (b) marketing, service, or other external factors. This classification serves as a heuristic guide, as variables in category (b) are more likely to be confounders, while those in category (a) typically represent direct components of user preferences rather than confounders. Following this preliminary classification, the LLM $\Phi$ conducts a thorough evaluation of each variable against the confounder criteria: a variable is identified as a confounder only if it both directly affects user-item interactions and indirectly affects user-item interactions by influencing user preferences. (iii) \emph{Output format control}, which specifies that the LLM $\Phi$ should produce a structured output containing all identified confounders, each with its description and the corresponding reasoning process justifying its identification as an observed confounder.

In our implementation, we initially employ zero-shot prompting using these three components without examples. After each extraction round, we store newly identified confounders, along with their descriptions and corresponding reasoning processes, into a confounder pool $\mathcal{C}_{pool}$ for subsequent analysis. Formally, at round $\tau$, this process can be represented as:
\begin{equation}
\mathcal{C}_{pool}^{\tau} := \mathcal{C}_{pool}^{\tau-1} \cup \Phi(\rho_{ext}, \mathcal{Z}_{MB}^{\leq \tau}),
\end{equation}
where $\mathcal{C}_{pool}^{0} = \emptyset$ initially, and $\Phi(\rho_{ext}, \mathcal{Z}_{MB}^{\leq \tau})$ represents the newly identified confounders at round $\tau$. Once $\mathcal{C}_{pool}$ contains at least one confounder, we transition to few-shot prompting for subsequent rounds, using both a positive example (identified confounder) and a negative example (non-confounder). This transition involves adding an example demonstration component to our prompt $\rho_{ext}$, positioned before the input data and context component. For the positive example, we select the first identified confounder from $\mathcal{C}_{pool}$. For the negative example, we randomly select a variable that was explicitly determined not to be a confounder. The pairing of these contrasting examples (few-shot) creates clearer decision boundaries for the LLM, allowing it to better distinguish between confounders and non-confounders based on their causal influence patterns, rather than relying solely on the confounder definition alone (zero-shot) or a positive example alone (one-shot). Each example includes the refined variable set, domain information, variable name, its description, and reasoning process that led to its determination as a confounder or non-confounder. In addition, we instruct the LLM $\Phi$ to avoid redundantly identifying confounders that are already present in the confounder pool $\mathcal{C}_{pool}$.

After completing all extraction iterations, we transform each confounder in $\mathcal{C}_{pool}$ into an embedding. First, we concatenate all textual information of the confounder (i.e., name, description, and reasoning process). Next, we encode this concatenated text using Sentence-BERT \cite{reimers2019sentence}, chosen for its efficiency, cost-effectiveness, and comparable performance to other embedding models (such as OpenAI's text embeddings) \cite{sun2024adaptive}. This encoding produces a 384-dimensional embedding, which we then reduce to $k$ dimensions using Principal Component Analysis (PCA), aligning with our user and item embedding dimensions. To minimize semantic redundancy and extract representative confounders, we apply K-means clustering to all the confounder embeddings and select $J^s$ cluster centroids to form our confounder subspace $\mathcal{C}^t$ \cite{zhu2025causal}. Similarly, we can obtain source domain confounder subspace $\mathcal{C}^s$. 

While the causal structure obtained by the FCI algorithm allows conventional confounder detection methods \cite{vanderweele2019principles, williamson2014introduction} to identify confounders, such methods operate purely on structural criteria and return only confounder names without semantic context required for effective deconfounding. By contrast, our LLM-based confounder extraction module operates at the semantic level by injecting domain information into prompts to guide domain-specific reasoning. This enables the identification and verification of confounders within appropriate domain contexts, such as single-domain confounders and cross-domain confounders. Moreover, the LLM produces structured outputs for each identified confounder, including detailed descriptions and justifications, which support the construction of semantically informed confounder embeddings for effective deconfounding, capabilities that conventional confounder detection methods cannot provide.

Overall, our LLM-guided confounder discovery module mitigates potential LLM hallucinations by integrating theoretical validation with iterative refinement. In the variable refinement stage, we systematically filter spurious variables through CI tests, discover causal structures via the FCI algorithm, and extract the Markov Blanket to ensure only causally relevant variables are retained. In the confounder extraction stage, we operate on this theoretically-grounded variable set rather than directly on unstructured text, employing iterative few-shot prompting to improve extraction accuracy. This integration ensures that our confounder extraction process is theoretically sound and retains only variables with true causal relationships to user interactions, unlike approaches that directly rely on LLM-based extraction which may include inaccuracies. It provides reliable candidate confounders for effective deconfounding in cross-domain OOD recommendation.

\subsubsection{\textbf{Generating Causal Feedback}}
In the previous step, we have extracted observed confounders from the refined variable set $\mathcal{Z}_{MB}^{\leq \tau}$. However, LLMs typically struggle to propose all relevant variables in a single iteration \cite{wei2022chain,qiao2023reasoning}. When the proposed variable set is incomplete, the subsequent extraction of observed confounders is inevitably limited as well. To address this issue, we develop a causal feedback mechanism to guide the LLM $\Phi$ in proposing additional relevant variables in subsequent iterations, thereby progressively improving both variable discovery and confounder extraction throughout our iterative pipeline.

To determine whether our refined variable set $\mathcal{Z}_{MB}^{\leq \tau}$ is sufficient for explaining the target variable $Y^t$, we employ conditional entropy analysis. Conditional entropy quantifies the reduction in uncertainty about $Y^t$ when conditioning on a given set of variables, providing a statistical measure of information completeness. If our refined variable set is incomplete, there exists a set of additional variables $\hat{\mathcal{Z}}$ that would further reduce the uncertainty about $Y^t$. We instruct the LLM $\Phi$ to propose new variables $\hat{z} \in \hat{\mathcal{Z}}$ that satisfy:
\begin{equation}
H_{\hat{R}^{\tau+1}}(Y^t|\mathcal{Z}_{MB}^{\leq \tau}) - H_{\hat{R}^{\tau+1}}(Y^t|\mathcal{Z}_{MB}^{\leq \tau}, \hat{z}(R^t)) > 0,
\label{inequality}
\end{equation}
where $H(\cdot \mid \cdot)$ represents conditional entropy measured on $\hat{R}^{\tau+1}$, which is a subset of reviews specifically selected for the $(\tau+1)$-th iteration. When Eq.(\ref{inequality}) holds, it indicates that the current variables are insufficient to fully explain $Y^t$, highlighting the need to discover additional causal variables. Our key insight for identifying these variables is to focus on reviews where the current variable set $\mathcal{Z}_{MB}^{\leq \tau}$ performs poorly in explaining $Y^t$, as they likely contain information about undiscovered causal variables. We express this selection criterion as $\hat{R}^{\tau+1} = \arg\max_{\hat{R} \subset R^t} H_{\hat{R}}(Y^t|\mathcal{Z}_{MB}^{\leq \tau})$. Following the method introduced in \cite{causalcoat2024}, we frame this as a classification problem to identify reviews where the current variables fail to provide adequate predictive power. In our implementation, we use K-means clustering to group reviews based on their representations derived from the current variable set $\mathcal{Z}_{MB}^{\leq \tau}$. After dividing the reviews into multiple clusters, we compute the conditional entropy for each cluster and select the one with the highest conditional entropy as $\hat{R}^{\tau+1}$. This approach effectively identifies the subset $\hat{R}^{\tau+1}$ where significant uncertainty about $Y^t$ remains despite conditioning on current variables, providing informative samples to guide the discovery of additional causal variables in next iteration.

Specifically, we utilize the causal feedback mechanism to generate inputs for the next iteration of variable proposal. This mechanism can be formalized as:
\begin{equation}
(\rho_{pro}^{\tau+1}, \hat{R}^{\tau+1}) = \xi(\mathcal{Z}_{MB}^{\leq \tau}, R^t, \rho_{pro}^\tau),
\end{equation}
where $\xi$ represents the causal feedback operation that: (1) selects the samples $\hat{R}^{\tau+1}$ from the highest conditional entropy cluster using the approach described earlier, and (2) generates an enhanced prompt $\rho_{pro}^{\tau+1}$ by modifying $\rho_{pro}^\tau$ to incorporate such samples and add feedback instructions. These instructions guide the LLM $\Phi$ to propose new causal variables not currently included in $\mathcal{Z}_{MB}^{\leq \tau}$, facilitating the construction of a more complete causal variable set in subsequent iterations.

\subsection{Causal Deconfounding and Cross-domain OOD Recommendation}
After obtaining the confounder subspaces $\mathcal{C}^s$ and $\mathcal{C}^t$ through LLM-guided confounder discovery, it becomes crucial to address the confounding biases introduced by the confounders within these subspaces. Without proper handling, such confounding biases can distort the inference of causal-invariant user preferences, thereby compromising the effectiveness of cross-domain OOD recommendation. To effectively mitigate these biases, we adopt the causal deconfounding approach proposed in \cite{zhu2025causal}, which leverages backdoor adjustment \cite{zhu2022deep} to estimate the direct causal effect from user preferences $E$ to interactions $Y$. This approach seamlessly integrates deconfounding into the recommendation training pipeline by modeling $P(Y|E,c)$ with an MLP, allowing us to estimate the causal effect as follows:
\begin{equation}
P(Y|do(E)) = {\mathbb{E}_c}[P(Y|E,c)] = {\mathbb{E}_c}[\varphi(\mathbf{E}_u^{*inv},{\mathbf{E}_v},\mathbf{c})],
\end{equation}
where $P$ denotes the probability and $\mathbb{E}$ denotes the expectation. $\varphi(\cdot)$ represents an MLP for predicting interaction probabilities \cite{he2017neural} and $c$ is the confounder selected from the corresponding confounder subspace $\mathcal{C}$. $\mathbf{E}_u^{*inv}$ denotes the comprehensive causal-invariant user preferences, and $\mathbf{E}_v$ represents the item embeddings. By implementing backdoor adjustment directly through the prediction process, this approach effectively blocks the backdoor paths from confounders to user-item interactions, enabling the inference of debiased causal-invariant user preferences that are critical for reliable cross-domain OOD recommendation.

Furthermore, inspired by the method introduced in \cite{zhang2023video}, we apply a confounder selection function to control the deconfounding process by appropriately weighting the confounders. Taking the target domain $D^t$ as an example, we formulate this confounder selection function as:
\begin{equation}
\psi (\mathbf{E}_u^{*tinv},\mathbf{E}_v^t,\mathbf{c}) = \frac{\exp (\mathbf{W}_u^t\mathbf{E}_u^{*tinv} \cdot \mathbf{W}_{uc}^t\mathbf{c})}{2\sum\nolimits_{\mathbf{c}'} \exp (\mathbf{W}_u^t\mathbf{E}_u^{*tinv} \cdot \mathbf{W}_{uc}^t\mathbf{c}')} + \frac{\exp (\mathbf{W}_v^t\mathbf{E}_v^t \cdot \mathbf{W}_{vc}^t\mathbf{c})}{2\sum\nolimits_{\mathbf{c}'} \exp (\mathbf{W}_v^t\mathbf{E}_v^t \cdot \mathbf{W}_{vc}^t\mathbf{c}')},
\end{equation}
where $\mathbf{W}_u^t$, $\mathbf{W}_{uc}^t$, $\mathbf{W}_v^t$, and $\mathbf{W}_{vc}^t$ denote learnable transformation matrices. $\cdot$ denotes dot product and $\mathbf{c}'$ represents any confounder from $\mathcal{C}^t$. Next, the expectation $\mathbb{E}_c[\varphi(\mathbf{E}_u^{*tinv},\mathbf{E}_v^t,\mathbf{c})]$ is expressed as:
\begin{equation}
\begin{split}
\mathbb{E}_c[\varphi(\mathbf{E}_u^{*tinv},\mathbf{E}_v^t,\mathbf{c})] = \varphi[\mathbf{W}_{fc}^t(\mathbf{E}_u^{*tinv}||\mathbf{E}_v^t||\sum\nolimits_c {p(c)\mathbf{c}} \psi(\mathbf{E}_u^{*tinv},\mathbf{E}_v^t,\mathbf{c}))],
\end{split}
\end{equation}
where $||$ denotes concatenation operation, and $\mathbf{W}_{fc}^t$ denotes a weight matrix of fully connected (FC) layer in the target domain. For practical purposes, the prior probability $p(c)$ is assumed as a uniform distribution. Moreover, $\boldsymbol{\Theta}_{in}^t={\mathbf{W}_{fc}^t}(\mathbf{E}_u^{*tinv}||\mathbf{E}_v^t||\sum\nolimits_c {p(c)\mathbf{c}} \psi (\mathbf{E}_u^{*tinv},\mathbf{E}_v^t,\mathbf{c})$ is input of MLP in the target domain. Likewise, we can obtain $\boldsymbol{\Theta}_{in}^s$ in the source domain as well. Furthermore, we obtain the predicted interaction $\hat y_{ij}^t$ between user $u_i$ and item $v_j$ within the target domain as follows:
\begin{equation}
\hat y_{ij}^t = \mathbf{\delta}_{out}(\mathbf{\delta}_l(...\mathbf{\delta}_2(\mathbf{\delta}_1(\boldsymbol{\Theta}_{in}))...)),
\end{equation}
where $\mathbf{\delta}_l^t$ represents the transformation operation at the $l$-th layer of MLP. The MLP comprises $l$ layers, with $\delta_{out}$ serving as the final output transformation.

In addition, we employ the cross-entropy loss to formulate the recommendation loss for the target domain as follows:
\begin{equation}
\mathcal{L}_{rec}^t = -\sum\limits_{y \in \mathcal{Y}^{t+} \cup \mathcal{Y}^{t-}} \left[ y \log(\hat{y}) + (1-y) \log(1-\hat{y}) \right],
\end{equation}
where $\hat{y}$ represents the predicted interaction probability and $y$ denotes the corresponding ground truth label. The set $\mathcal{Y}^{t+}$ contains all observed positive interactions in the target domain, while $\mathcal{Y}^{t-}$ consists of negative samples randomly selected from unobserved interactions to mitigate overfitting. Likewise, we can obtain $\hat y_{ij}^s$ and $\mathcal{L}_{rec}^s$ in source domain.

Finally, we formulate the total loss function as follows:
\begin{equation}
\mathcal{L}^{total} = \mathcal{L}_{rec}^t + \beta_1 \mathcal{L}_{rec}^s + \beta_2 \mathcal{L}^{\text{cau}} + \beta_3 \mathcal{L}^{\text{dom}} + \beta_4 \|\mathbf{\Omega}\|_2,
\end{equation}
where $\mathbf{\Omega}$ represents the set of all parameters that are optimized during training. The hyperparameters $\beta_1$ through $\beta_4$ control the contribution of each term.

\subsection{Time Complexity Analysis}
\label{time_complexity}
In this section, we analyze the time complexity of our proposed CICDOR framework. To standardize our analysis, we assume consistent parameters in CICDOR: the number of network layers $L$ and embedding dimensions $k$. Below, we assess the time complexity of each of four functional modules individually.

\noindent \textbf{(1) User Preference Disentanglement:} This module consists of two main computational phases. First, during the embedding construction and graph propagation phase, we transform users and items into enriched embeddings. Since the text embedding is precomputed as a one-time preprocessing step, its time complexity can be excluded from the analysis. Assuming the heterogeneous graph contains $(m+n)$ nodes with an average of $\bar{N}$ neighbors per node, the time complexity of the $L$-layer GCN operation is $O(L(m+n)\bar{N}k)$, where $m$ and $n$ denote the number of users and items, respectively. Second, in the user preference disentanglement phase, we employ three MLP-based encoders along with a domain discriminator. The time complexity of this phase is dominated by the $L$-layer MLP operations, approximately $O(Lmk^2)$. Combining both phases and noting that typically $\bar{N} \ll (m+n)$, the overall time complexity of this module simplifies to $O(L(m+n)k + Lmk^2)$.

\noindent \textbf{(2) Dual-level Causal Preference Learning:} This module learns casual structure to infer causal-invariant user preferences at both domain-specific and domain-shared levels. The time complexity analysis spans both training and inference phases. In the training phase, learning the causal structure through adjacency matrices involves (i) matrix multiplication for reconstruction loss computation with complexity $O(mk^2)$, and (ii) acyclicity constraint calculation with complexity $O(k^3)$. Additional regularization terms (e.g., sparsity, path constraints) contribute $O(k^2)$ complexity but are dominated by the higher-order terms. In the inference phase, inferring causal-invariant user preferences requires similar matrix multiplication operations with complexity $O(mk^2)$, while the attention-based fusion of preferences from both levels adds $O(mk)$ complexity. Thus, the overall time complexity of this module simplifies to $O(mk^2 + k^3)$.

\noindent \textbf{(3) LLM-guided Confounder Discovery:} This module identifies causal variables and confounders through iterative processing across five key steps. (i) For variable proposal, the LLM processes review samples to generate candidate causal variables with complexity $O(\mathcal{T}_{\Phi}(|\bar{x}|))$, where $\mathcal{T}_{\Phi}(\cdot)$ represents the LLM inference cost, and $|\bar{x}|$ denotes the average input length. (ii) For review annotation, each review is annotated based on the criteria of newly proposed variables, requiring complexity $O(|R||\mathcal{Z}|\mathcal{T}_{\Phi}(|\bar{x}|))$, where $|R|$ denotes the number of reviews and $|\mathcal{Z}|$ denotes the number of extracted variables. (iii) For variable refinement, CI test requires correlation matrix computation with complexity $O(|R||\mathcal{Z}|^2)$, the FCI algorithm contributes $O(|\mathcal{Z}|^4)$, and Markov Blanket extraction adds $O(|\mathcal{Z}|^2)$, resulting in a combined complexity of $O(|R||\mathcal{Z}|^2 + |\mathcal{Z}|^4)$. (iv) For confounder extraction, the LLM processes the refined variable set in one call with complexity $O(\mathcal{T}_{\Phi}(|\bar{x}|))$. (v) For causal feedback, K-means clustering and conditional entropy calculation require $O(|R||\mathcal{Z}|JI)$ complexity, where $J$ is the number of cluster centroids and $I$ is the number of clustering iterations. Additionally, after all iterations, we process extracted confounders through embedding generation, dimensionality reduction, and clustering with complexity $O(|\mathcal{C}|(|\bar{x}| + k^2 + kJI))$, where $|\mathcal{C}|$ represents the number of extracted confounders. Considering all steps across $\tau$ iterations, and noting that $|\mathcal{Z}| \ll |R|$, $\tau \ll |R||\mathcal{Z}|$, and computation costs are dominated by LLM inference, the overall time complexity of this module simplifies to $O(|R||\mathcal{Z}|\mathcal{T}_{\Phi}(|\bar{x}|) + \tau|R||\mathcal{Z}|^2)$.

\noindent \textbf{(4) Causal Deconfounding and Cross-domain OOD Recommendation:} This module employs backdoor adjustment to mitigate confounding biases and estimate the direct causal effects of user preferences on interactions. Computing the selection weights for each confounder across all user-item pairs has a complexity of $O(mnJk)$. Subsequently, the process of concatenating user, item, and weighted confounder embeddings and passing them through an $L$-layer MLP contributes an additional complexity of $O(Lmnk^2)$. Consequently, the overall time complexity of this module is $O(mnk(J + Lk))$.

Overall, the time complexity of our CICDOR framework encompasses two distinct phases: the LLM-guided confounder discovery with complexity $O(|R||\mathcal{Z}|\mathcal{T}_{\Phi}(|\bar{x}|))$, and the recommendation model training (consisting of the remaining three modules) with complexity $O(mnLk^2)$. These expressions represent the simplified dominant terms from our detailed module-by-module analysis. The LLM-guided confounder discovery phase theoretically scales with the number of processed reviews, extracted variables, and LLM inference cost; however, in common practice, reviews are typically sampled at a computationally tractable scale to balance between representativeness and efficiency. Meanwhile, the recommendation training phase scales linearly with the user-item interaction space and quadratically with embedding dimensions, which typically becomes the computational bottleneck in large-scale recommendation scenarios.

\section{Experiments and Analysis}
To validate the efficacy of our proposed CICDOR framework and its different modules, we conduct extensive experiments on two real-world datasets to answer the following research questions:

\begin{itemize}[leftmargin=*]
    \item \noindent \textbf{RQ1.} How does our CICDOR perform when evaluated across existing state-of-the-art approaches (see Section \ref{sec:V-B})?
    \item \noindent \textbf{RQ2.} How do various modules within our framework, namely, dual-level causal preference learning and LLM-guided confounder discovery, influence the OOD recommendation performance in the target domain? (see Section \ref{sec:V-C})?
    \item \noindent \textbf{RQ3.} How does our CICDOR's performance vary across different degrees of distribution shift and different overlapping user ratios (see Section \ref{sec:V-D})?
    \item \noindent \textbf{RQ4.} How does adjusting different hyperparameters influence the performance of our CICDOR (see Section \ref{sec:V-E})?
    \item \noindent \textbf{RQ5.} How does CICDOR's performance vary under annotation noise and prompt perturbations in the LLM-guided confounder discovery module (see Section \ref{sec:V-F})?
    \item \noindent \textbf{RQ6.} What are the underlying causal mechanisms learned by our CICDOR (see Section \ref{sec:V-G})?

\end{itemize}

\begin{table*}[t]
\caption{Statistics of the datasets.}
\label{tab:dataset_statistics}
\vspace{-0.1in}
\begin{tabular}{cccccc}
\hline
Datasets                   & Domains     & \#Users & \#Items & \#Interactions & Density \\ \hline
\multirow{2}{*}{Douban} & Movie (source) & 2106    & 9555    & 907219         & 4.508\% \\
                        & Book (target)  & 2106    & 6777    & 95974          & 0.672\% \\ \hline
\multirow{2}{*}{Douban} & Movie (source) & 1666    & 9555    & 781288         & 4.908\% \\
                        & Music (target) & 1666    & 5567    & 69681          & 0.751\% \\ \hline
\multirow{2}{*}{Amazon} & Electronics (source)  & 15761   & 51447   & 224689         & 0.027\% \\
                        & Clothing (target) & 15761   & 48781   & 133609         & 0.017\% \\ \hline
\end{tabular}
\vspace{-0.1in}
\end{table*}

\subsection{Experimental Settings}
\subsubsection{\textbf{Experimental Datasets and Task Settings}}
We conduct experiments on two widely used CDR datasets: Douban \cite{zhu2020graphical} and Amazon \cite{cao2022disencdr}. For clarity, we refer to the specific domains as Douban-Movie, Douban-Book, Douban-Music, Amazon-Elec, and Amazon-Cloth throughout this paper. Based on these domains, we construct three source-target domain pairs: (1) Douban-Movie $\rightarrow$ Douban-Book, (2) Douban-Movie $\rightarrow$ Douban-Music, and (3) Amazon-Elec $\rightarrow$ Amazon-Cloth, where the first domain in each pair serves as the source domain and the second as the target domain. Table~\ref{tab:dataset_statistics} presents the statistics of these domain pairs. Both datasets contain user-generated ratings and textual reviews. As the original ratings are explicit feedback, we convert them into implicit feedback by treating interactions with ratings of 4 or higher as positive instances. Moreover, we extract overlapping users in both the source and target domains of each pair, thereby constructing a full user overlap scenario.

For all datasets, we first evaluate the models under an IID setting, where the training and testing data are randomly sampled from the same distribution. Following the existing works \cite{he2022causpref,li2024cross}, we then conduct experiments under three OOD settings in the target domain:
\begin{itemize}
    \item \textbf{User Degree Shift (OOD \#1 and OOD \#2):} Users with different interaction degrees often exhibit distinct behavior patterns. In this setting, the training set is randomly sampled from the entire dataset, covering users with varying interaction degrees, while the test set consists exclusively of low-degree users (\textbf{OOD \#1}) or high-degree users (\textbf{OOD \#2}). These variants create distribution shifts in user interaction degrees, challenging the model to generalize from the overall user population to specific user subsets with extreme interaction degrees.
    \item \textbf{Region Shift (OOD \#3):} User preferences often differ across different regions. In this setting, the training set contains randomly sampled users from all regions, while the test set is composed of users from Beijing. This simulates a real-world scenario where regional distribution shift occurs between training and testing data.
\end{itemize}
For all experiments, we adopt an 8:1:1 split ratio for training, validation, and testing data. The user degree shift setting is applied to both Douban and Amazon datasets, while the region shift setting is applied only to the Douban dataset because only this dataset contains region information.

\subsubsection{\textbf{Implementation Details}}
For the hyperparameters in our causal structure learning component, we adopt the weights for different terms in the loss function as used in \cite{li2024cross}. For the overall model training, we employ a negative sampling strategy with a 1:3 ratio of positive to negative samples. In addition, we set the trade-off parameters $\beta_1 = 1.0$, $\beta_3 = 1$, $\beta_4 = 0.00001$ and $\gamma = 0.5$ for the corresponding loss terms, while setting the number of cluster centroids $J=J^s=J^t$ to 10 and the dual-level causal loss weight $\beta_2$ to 0.5 as our default configuration. 

For the prediction network, we implement a three-layer MLP to process the concatenated user, item, and confounder embeddings. Following \cite{zhu2025causal}, we set the dimension after the FC layer $k_{in}$ as 128 and the output dimension $k_{out}$ as 8, after exploring values in ranges \{64, 128\} and \{8, 16\}, respectively. For the LLM-guided confounder discovery module, we sample 1000 users from each dataset and collect up to 5 reviews per user as input, utilizing gpt-4o-mini as the LLM with API calls determined at zero temperature to ensure reproducible outputs.

To find out the optimal hyperparameters, we employ Optuna\footnote{https://optuna.org/} with 50 trials for each model. The search space includes learning rate in \{$10^{-4}$, $10^{-3}$, $10^{-2}$\}, batch size in \{128, 256, 512\}, and embedding dimension $k$ in \{32, 64, 128\}. Based on these optimization results, we use the Adam optimizer \cite{kingma2014adam} for all models. In Section \ref{sec:V-E}, we present an extensive parameter sensitivity analysis to examine how variations in the number of cluster centroids $J$ in \{2, 5, 10, 20, 50\} and the dual-level causal loss weight $\beta_2$ in \{0.1, 0.25, 0.5, 0.75, 1.0\} affect the overall model performance.

\subsubsection{\textbf{Model Training}}
\label{model_training}
Our CICDOR framework adopts a two-phase training strategy designed to ensure stable convergence and accurate causal structure learning. This design is based on the principle that learning reliable causal relationships requires a foundation of high-quality user representations, as attempting to learn causal structures before the representations are well-formed would result in inaccurate or unstable causal structures.

In the first phase, we train the model for 60 epochs\footnote{We select the number of training epochs from the range $\{20, 40, 60, 80\}$ in each phase.} to learn user attributes, domain-shared and domain-specific user preferences, while simultaneously performing deconfounding using fixed confounder embeddings extracted via our LLM-guided confounder discovery module. This phase generates debiased domain-shared and domain-specific user preferences, forming debiased comprehensive user preferences.

In the second phase, we continue training for an additional 40 epochs and introduce the dual-level causal loss in Eq.~(\ref{eq8}), which guides the model to learn dual-level causal structures. Specifically, this phase identifies the causal relationships between user attributes and the debiased domain-shared and domain-specific user preferences, respectively. The resulting causal structures capture the invariant causal mechanisms and enable the inference of causal-invariant user preferences at both the domain-shared and domain-specific levels. These causal-invariant user preferences maintain consistency across distribution shifts, significantly improving OOD recommendation performance. For fair comparison, all baseline methods are also trained for 100 epochs to ensure complete convergence.

\subsubsection{\textbf{Evaluation Metrics}}
\label{evaluation_metrics}
For our evaluation, we adopt two widely used metrics in the evaluation of RSs: Hit Ratio (HR) and Normalized Discounted Cumulative Gain (NDCG) \cite{li2024cross}. During testing, for each positive user-item interaction, we randomly sample 99 items that the user has not interacted with as negative samples. The model then ranks this candidate set of 100 items (1 positive + 99 negative), and we evaluate its performance based on the position of the positive item within this ranked list. Throughout our experimental analysis, we focus specifically on the top-10 positions in the ranking results. To ensure the reliability of our findings, all experiments are conducted five times, and we report the average performance across these runs.

\subsubsection{\textbf{Comparison Methods}}
We compare our proposed CICDOR with twelve representative and state-of-the-art baseline models, which can be categorized into four groups: (I) Disentanglement-based OOD Recommendation, (II) Causality-based OOD Recommendation, (III) Single-target Cross-domain Recommendation (CDR), and (IV) Adaptation-based OOD Recommendation. Given that our work focuses on the novel setting of cross-domain OOD recommendation, which essentially addresses single-target CDR in OOD environments, we select representative and state-of-the-art methods from both single-target CDR and OOD recommendation literature as our baselines \cite{zhu2024active}. Here, single-target CDR \cite{zhu2019dtcdr} refers to the paradigm of transferring knowledge from data-richer source domain to improve recommendation performance in the data-sparser target domain. We do not include dual-target CDR methods \cite{ijcai2021p639} as baselines, because they aim to improve recommendation performance in both domains simultaneously, which differs from our setting.

Although several approaches explore invariance and generalization principles, we do not include them in the baselines as they address different settings from our cross-domain OOD recommendation. These excluded approaches target domain generalization \cite{zhang2023connecting,zhang2024transferring}, sequential recommendation \cite{zhang2024disentangled,xu2024rethinking}, item cold-start recommendation \cite{wang2023equivariant}, and recommendation with noisy interaction data \cite{zhao2025distributionally}. Detailed descriptions of selected baseline models are provided below.

\begin{itemize}[leftmargin=*]
\item \textbf{DICE} \cite{zheng2021disentangling} (I) decouples user interest and conformity by modeling their causal generation process, learning disentangled embeddings through cause-specific sampling guided by the collider effect.
\item \textbf{DCCL} \cite{zhao2023disentangled} (I) is a model-agnostic framework that addresses OOD problems and data sparsity by disentangling interest and conformity through contrastive learning with cause-specific sample augmentation.
\item \textbf{CausPref} \cite{he2022causpref} (II) learns invariant user preferences from implicit feedback via causal structure learning and enhances robustness to distribution shift through anti-preference negative sampling.
\item \textbf{COR} \cite{wang2022causal} (II) formulates user feature shift as an intervention and performs causal inference with a tailored VAE to estimate post-intervention interaction probabilities for robust OOD recommendation.
\item \textbf{InvCF} \cite{zhang2023invariant} (II) identifies causally invariant preference representations by disentangling them from popularity semantics, enabling consistent generalization in real-world scenarios with shifting item popularity.
\item \textbf{PopGo} \cite{zhang2024robust} (II) improves OOD generalization by mitigating interaction-level popularity shortcuts, using a learned shortcut model to adjust predictions and emphasize true user preferences over spurious popularity shortcuts. 
\item \textbf{CausalDiffRec} \cite{zhao2025graph} (II) enhances OOD recommendation by eliminating environmental confounders through backdoor adjustment and learning environment-invariant graph representations via a causal diffusion process.
\item \textbf{PTUPCDR} \cite{zhu2022personalized} (III) generates personalized bridge functions through a task-optimized meta network, facilitating stable and personalized preference transfer from the source to the target domain.
\item \textbf{CUT} \cite{li2024aiming} (III) employs a two-phase training strategy that first captures user similarities in the target domain and then transfers source-domain information selectively, leveraging a user transformation module and contrastive learning to avoid relationship distortion.
\item \textbf{CDCOR} \cite{li2024cross} (III) improves OOD recommendation by transferring cross-domain knowledge, leveraging a domain adversarial network to extract shared user preferences and a causal structure learner to model invariant relationships under distribution shifts. Notably, \textbf{CDCOR} represents a strong single-level causal baseline that focuses on domain-shared causal structure, rather than explicitly modeling domain-specific causal structures related to SDDS.
\item \textbf{DR-GNN} \cite{wang2024distributionally} (IV) integrates distributionally robust optimization (DRO) into GNN-based recommendation by treating GNN as a smoothing regularizer and injecting small perturbations into sparse neighbor distributions to enhance robustness against distribution shifts.
\item \textbf{DT3OR} \cite{yang2025dual} (IV) introduces a dual test-time training strategy for OOD recommendation, adapting models to distribution shifts by learning invariant user preferences and variant user/item features through self-distillation and contrastive learning.
\end{itemize}

\begin{table}[ht]
\caption{Quantitative assessment of CICDOR against representative and state-of-the-art baseline models across two datasets under IID and OOD settings (OOD \#1), measured by HR@10 and NDCG@10 metrics. The best results appear in bold text, with the best baseline results identified by underlining. * represents $p < 0.05$ when CICDOR is compared to the best baseline in paired t-test \protect\cite{zhu2023domain}.}
\vspace{-0.1in}
\label{IID_comparison}
\resizebox{\textwidth}{!}{%
\begin{tabular}{cc|cccccccc|cccc}
\hline
\multicolumn{2}{c|}{\textbf{Dataset}} & \multicolumn{8}{c|}{\textbf{Douban}} & \multicolumn{4}{c}{\textbf{Amazon}} \\ \hline
\multicolumn{2}{c|}{\textbf{Domain: Source $\rightarrow$ Target}} & \multicolumn{4}{c|}{\textbf{Movie $\rightarrow$ Book}} & \multicolumn{4}{c|}{\textbf{Movie $\rightarrow$ Music}} & \multicolumn{4}{c}{\textbf{Elec $\rightarrow$ Cloth}} \\ \hline
\multicolumn{2}{c|}{Setting} & \multicolumn{2}{c|}{IID} & \multicolumn{2}{c|}{OOD \#1} & \multicolumn{2}{c|}{IID} & \multicolumn{2}{c|}{OOD \#1} & \multicolumn{2}{c|}{IID} & \multicolumn{2}{c}{OOD \#1} \\ \hline
\multicolumn{2}{c|}{Metric} & HR & \multicolumn{1}{c|}{NDCG} & HR & \multicolumn{1}{c|}{NDCG} & HR & \multicolumn{1}{c|}{NDCG} & HR & NDCG & HR & \multicolumn{1}{c|}{NDCG} & HR & NDCG \\ \hline
\multicolumn{1}{c|}{\multirow{2}{*}{\textbf{I}}} & DICE & 0.3698 & \multicolumn{1}{c|}{0.2089} & 0.3043 & \multicolumn{1}{c|}{0.1698} & 0.3237 & \multicolumn{1}{c|}{0.1755} & 0.2753 & 0.1485 & 0.6439 & \multicolumn{1}{c|}{0.3741} & 0.5220 & 0.2794 \\
\multicolumn{1}{c|}{} & DCCL & 0.3870 & \multicolumn{1}{c|}{0.2195} & 0.3462 & \multicolumn{1}{c|}{0.1837} & 0.3431 & \multicolumn{1}{c|}{0.1938} & 0.3097 & 0.1704 & 0.6627 & \multicolumn{1}{c|}{0.3994} & 0.5811 & 0.3267 \\ \hline
\multicolumn{1}{c|}{\multirow{5}{*}{\textbf{II}}} & CausPref & 0.3334 & \multicolumn{1}{c|}{0.1901} & 0.2856 & \multicolumn{1}{c|}{0.1553} & 0.2997 & \multicolumn{1}{c|}{0.1654} & 0.2502 & 0.1365 & 0.5723 & \multicolumn{1}{c|}{0.3222} & 0.4776 & 0.2633 \\
\multicolumn{1}{c|}{} & COR & 0.3646 & \multicolumn{1}{c|}{0.2062} & 0.3015 & \multicolumn{1}{c|}{0.1679} & 0.3165 & \multicolumn{1}{c|}{0.1742} & 0.2698 & 0.1440 & 0.6195 & \multicolumn{1}{c|}{0.3498} & 0.5035 & 0.2704 \\
\multicolumn{1}{c|}{} & InvCF & 0.3739 & \multicolumn{1}{c|}{0.2125} & 0.3194 & \multicolumn{1}{c|}{0.1757} & 0.3369 & \multicolumn{1}{c|}{0.1867} & 0.2859 & 0.1553 & 0.6374 & \multicolumn{1}{c|}{0.3705} & 0.5236 & 0.2798 \\
\multicolumn{1}{c|}{} & PopGo & 0.3458 & \multicolumn{1}{c|}{0.1963} & 0.2918 & \multicolumn{1}{c|}{0.1606} & 0.3058 & \multicolumn{1}{c|}{0.1732} & 0.2596 & 0.1411 & 0.5956 & \multicolumn{1}{c|}{0.3389} & 0.4882 & 0.2690 \\
\multicolumn{1}{c|}{} & CausalDiffRec & 0.3857 & \multicolumn{1}{c|}{0.2184} & 0.3533 & \multicolumn{1}{c|}{0.1871} & 0.3456 & \multicolumn{1}{c|}{0.1941} & 0.3152 & 0.1738 & 0.6657 & \multicolumn{1}{c|}{0.4007} & 0.5941 & 0.3323 \\ \hline
\multicolumn{1}{c|}{\multirow{3}{*}{\textbf{III}}} & PTUPCDR & 0.3472 & \multicolumn{1}{c|}{0.1976} & 0.3025 & \multicolumn{1}{c|}{0.1684} & 0.2983 & \multicolumn{1}{c|}{0.1644} & 0.2665 & 0.1427 & 0.5821 & \multicolumn{1}{c|}{0.3270} & 0.4974 & 0.2715 \\
\multicolumn{1}{c|}{} & CUT & 0.3915 & \multicolumn{1}{c|}{0.2241} & 0.2832 & \multicolumn{1}{c|}{0.1547} & 0.3255 & \multicolumn{1}{c|}{0.1762} & 0.2544 & 0.1371 & 0.6618 & \multicolumn{1}{c|}{0.3992} & 0.4797 & 0.2642 \\
\multicolumn{1}{c|}{} & CDCOR & 0.3821 & \multicolumn{1}{c|}{0.2172} & 0.3486 & \multicolumn{1}{c|}{0.1849} & 0.3413 & \multicolumn{1}{c|}{0.1919} & 0.3167 & 0.1726 & 0.6674 & \multicolumn{1}{c|}{0.4013} & 0.5926 & 0.3338 \\ \hline
\multicolumn{1}{c|}{\multirow{2}{*}{\textbf{IV}}} & DR-GNN & 0.3709 & \multicolumn{1}{c|}{0.2054} & 0.3418 & \multicolumn{1}{c|}{0.1831} & 0.3292 & \multicolumn{1}{c|}{0.1868} & 0.3032 & 0.1676 & 0.6505 & \multicolumn{1}{c|}{0.3786} & 0.5679 & 0.3193 \\
\multicolumn{1}{c|}{} & DT3OR & {\ul 0.3984} & \multicolumn{1}{c|}{{\ul 0.2289}} & {\ul 0.3660} & \multicolumn{1}{c|}{{\ul 0.1997}} & {\ul 0.3561} & \multicolumn{1}{c|}{{\ul 0.1957}} & {\ul 0.3297} & {\ul 0.1758} & {\ul 0.6729} & \multicolumn{1}{c|}{{\ul 0.4041}} & {\ul 0.6127} & {\ul 0.3465} \\ \hline
\multicolumn{1}{c|}{\textbf{Our}} & CICDOR & \textbf{0.4136*} & \multicolumn{1}{c|}{\textbf{0.2450*}} & \textbf{0.3923*} & \multicolumn{1}{c|}{\textbf{0.2244*}} & \textbf{0.3706*} & \multicolumn{1}{c|}{\textbf{0.2064*}} & \textbf{0.3525*} & \textbf{0.1949*} & \textbf{0.6938*} & \multicolumn{1}{c|}{\textbf{0.4216*}} & \textbf{0.6448*} & \textbf{0.3732*} \\ \hline
\multicolumn{2}{c|}{\textbf{\begin{tabular}[c]{@{}c@{}}Improvement (CICDOR \\ vs. best baselines)\end{tabular}}} & 3.82\% & \multicolumn{1}{c|}{7.03\%} & 7.19\% & \multicolumn{1}{c|}{12.37\%} & 4.07\% & \multicolumn{1}{c|}{5.47\%} & 6.92\% & 10.86\% & 3.11\% & \multicolumn{1}{c|}{4.33\%} & 5.24\% & 7.71\% \\ \hline
\end{tabular}%
}
\vspace{-0.1in}
\end{table}

\subsection{Performance Comparison (for RQ1)}
\label{sec:V-B}
To evaluate OOD robustness, Table \ref{IID_comparison} compares CICDOR against baseline models across two datasets under both IID and the challenging low-degree user settings (OOD \#1), using HR@10 and NDCG@10 as metrics. Moreover, Table \ref{baseline_comparison} presents the performance evaluation of CICDOR\footnote{Due to page constraints, we only report results with embedding dimension $k=64$, though CICDOR consistently outperforms baseline models across other embedding dimensions as well.} against various baseline models across two datasets under the other two OOD settings, i.e., the high-degree user setting (OOD \#2) and the region shift setting (OOD \#3). It should be noted that while Single-target CDR models are trained on both domains, we only report their performance in the target domain, as these models are specifically designed to improve recommendation performance in the target domain. From Table \ref{IID_comparison} and Table \ref{baseline_comparison}, we can observe that:

\begin{itemize}[leftmargin=*]
\item [(1)] Table \ref{IID_comparison} reveals that almost all baseline methods experience noticeable performance degradation when moving from IID to OOD settings (OOD \#1), indicating their sensitivity to distribution shifts. This effect is particularly pronounced for methods such as CUT \cite{li2024aiming}, which achieve strong performance under IID settings but suffer substantial declines under OOD settings. On average, baseline methods exhibit performance drops of 13.94\% in HR@10 and 17.34\% in NDCG@10. In contrast, CICDOR demonstrates the smallest average degradation, with decreases of only 5.70\% in HR@10 and 8.49\% in NDCG@10, highlighting its superior 
robustness under OOD settings\footnote{CICDOR's superior robustness is also observed across other OOD settings (OOD \#2 and OOD \#3).} in CDR.

\item[(2)] Our CICDOR demonstrates superior performance compared to Disentanglement-based OOD Recommendation baselines across various OOD settings, achieving average improvements of 18.89\% and 23.62\% w.r.t. HR@10 and NDCG@10, respectively. While DICE \cite{zheng2021disentangling} and DCCL \cite{zhao2023disentangled} employ factor disentanglement techniques, they are limited to handling only specific types of distribution shifts. In contrast, our CICDOR can effectively capture the invariant causal structure underlying user preferences. This enables us to infer causal-invariant user preferences for OOD recommendation, resulting in better performance across various types of distribution shifts;

\begin{table}[ht]
\caption{Quantitative assessment of CICDOR against representative and state-of-the-art baseline models across two datasets under different OOD settings (OOD \#2 and OOD \#3), measured by HR@10 and NDCG@10 metrics. The best results appear in bold text, with the best baseline results identified by underlining. * represents $p < 0.05$ when CICDOR is compared to the best baseline in paired t-test \protect\cite{zhu2023domain}.}
\vspace{-0.1in}
\label{baseline_comparison}
\resizebox{\textwidth}{!}{%
\begin{tabular}{cc|cccccccc|cc}
\hline
\multicolumn{2}{c|}{\textbf{Dataset}} & \multicolumn{8}{c|}{\textbf{Douban}} & \multicolumn{2}{c}{\textbf{Amazon}} \\ \hline
\multicolumn{2}{c|}{\textbf{Domain: Source $\rightarrow$ Target}} & \multicolumn{4}{c|}{\textbf{Movie $\rightarrow$ Book}} & \multicolumn{4}{c|}{\textbf{Movie $\rightarrow$ Music}} & \multicolumn{2}{c}{\textbf{Elec $\rightarrow$ Cloth}} \\ \hline
\multicolumn{2}{c|}{Setting} & \multicolumn{2}{c|}{OOD \#2} & \multicolumn{2}{c|}{OOD \#3} & \multicolumn{2}{c|}{OOD \#2} & \multicolumn{2}{c|}{OOD \#3} & \multicolumn{2}{c}{OOD \#2} \\ \hline
\multicolumn{2}{c|}{Metric} & HR & \multicolumn{1}{c|}{NDCG} & HR & \multicolumn{1}{c|}{NDCG} & HR & \multicolumn{1}{c|}{NDCG} & HR & NDCG & HR & NDCG \\ \hline
\multicolumn{1}{c|}{\multirow{2}{*}{\textbf{\begin{tabular}[c]{@{}c@{}}Disentanglement-based \\ OOD Recommendation\end{tabular}}}} & DICE & 0.3146 & \multicolumn{1}{c|}{0.1735} & 0.3406 & \multicolumn{1}{c|}{0.1827} & 0.2825 & \multicolumn{1}{c|}{0.1519} & 0.3047 & 0.1729 & 0.5311 & 0.2946 \\
\multicolumn{1}{c|}{} & DCCL & 0.3524 & \multicolumn{1}{c|}{0.1868} & 0.3812 & \multicolumn{1}{c|}{0.2165} & 0.3172 & \multicolumn{1}{c|}{0.1723} & 0.3438 & 0.1842 & 0.5948 & 0.3385 \\ \hline
\multicolumn{1}{c|}{\multirow{5}{*}{\textbf{\begin{tabular}[c]{@{}c@{}}Causality-based \\ OOD Recommendation\end{tabular}}}} & CausPref & 0.2911 & \multicolumn{1}{c|}{0.1604} & 0.3164 & \multicolumn{1}{c|}{0.1781} & 0.2601 & \multicolumn{1}{c|}{0.1416} & 0.2840 & 0.1544 & 0.4873 & 0.2682 \\
\multicolumn{1}{c|}{} & COR & 0.3115 & \multicolumn{1}{c|}{0.1716} & 0.3423 & \multicolumn{1}{c|}{0.1839} & 0.2758 & \multicolumn{1}{c|}{0.1487} & 0.2996 & 0.1655 & 0.5184 & 0.2778 \\
\multicolumn{1}{c|}{} & InvCF & 0.3253 & \multicolumn{1}{c|}{0.1761} & 0.3610 & \multicolumn{1}{c|}{0.1987} & 0.2905 & \multicolumn{1}{c|}{0.1589} & 0.3177 & 0.1788 & 0.5441 & 0.3017 \\
\multicolumn{1}{c|}{} & PopGo & 0.2988 & \multicolumn{1}{c|}{0.1652} & 0.3197 & \multicolumn{1}{c|}{0.1726} & 0.2682 & \multicolumn{1}{c|}{0.1434} & 0.2913 & 0.1606 & 0.5022 & 0.2755 \\
\multicolumn{1}{c|}{} & CausalDiffRec & 0.3595 & \multicolumn{1}{c|}{0.1981} & 0.3919 & \multicolumn{1}{c|}{0.2242} & 0.3246 & \multicolumn{1}{c|}{0.1758} & 0.3581 & 0.1968 & 0.6089 & 0.3394 \\ \hline
\multicolumn{1}{c|}{\multirow{3}{*}{\textbf{\begin{tabular}[c]{@{}c@{}}Single-target Cross-\\ domain Recommendation\end{tabular}}}} & PTUPCDR & 0.3108 & \multicolumn{1}{c|}{0.1714} & 0.3374 & \multicolumn{1}{c|}{0.1791} & 0.2721 & \multicolumn{1}{c|}{0.1473} & 0.2892 & 0.1574 & 0.5107 & 0.2763 \\
\multicolumn{1}{c|}{} & CUT & 0.2967 & \multicolumn{1}{c|}{0.1642} & 0.3248 & \multicolumn{1}{c|}{0.1757} & 0.2653 & \multicolumn{1}{c|}{0.1422} & 0.2865 & 0.1557 & 0.4965 & 0.2726 \\
\multicolumn{1}{c|}{} & CDCOR & 0.3586 & \multicolumn{1}{c|}{0.1974} & 0.3885 & \multicolumn{1}{c|}{0.2196} & 0.3217 & \multicolumn{1}{c|}{0.1745} & 0.3543 & 0.1956 & 0.6044 & 0.3381 \\ \hline
\multicolumn{1}{c|}{\multirow{2}{*}{\textbf{\begin{tabular}[c]{@{}c@{}}Adaptation-based \\ OOD Recommendation\end{tabular}}}} & DR-GNN & 0.3479 & \multicolumn{1}{c|}{0.1845} & 0.3803 & \multicolumn{1}{c|}{0.2162} & 0.3089 & \multicolumn{1}{c|}{0.1704} & 0.3348 & 0.1785 & 0.5792 & 0.3258 \\
\multicolumn{1}{c|}{} & DT3OR & {\ul 0.3713} & \multicolumn{1}{c|}{{\ul 0.2068}} & {\ul 0.4071} & \multicolumn{1}{c|}{{\ul 0.2289}} & {\ul 0.3336} & \multicolumn{1}{c|}{{\ul 0.1778}} & {\ul 0.3626} & {\ul 0.1993} & {\ul 0.6239} & {\ul 0.3570} \\ \hline
\multicolumn{1}{c|}{\multirow{8}{*}{\textbf{\begin{tabular}[c]{@{}c@{}}Our Model and \\ its Variants\end{tabular}}}} & CICDOR & \textbf{0.3951*} & \multicolumn{1}{c|}{\textbf{0.2264*}} & \textbf{0.4298*} & \multicolumn{1}{c|}{\textbf{0.2493*}} & \textbf{0.3564*} & \multicolumn{1}{c|}{\textbf{0.1961*}} & \textbf{0.3882*} & \textbf{0.2194*} & \textbf{0.6584*} & \textbf{0.3867*} \\
\multicolumn{1}{c|}{} & w/o dual-level & 0.3368 & \multicolumn{1}{c|}{0.1786} & 0.3680 & \multicolumn{1}{c|}{0.2051} & 0.3015 & \multicolumn{1}{c|}{0.1663} & 0.3280 & 0.1765 & 0.5648 & 0.3185 \\
\multicolumn{1}{c|}{} & w/o specific-level & 0.3675 & \multicolumn{1}{c|}{0.2003} & 0.3994 & \multicolumn{1}{c|}{0.2272} & 0.3326 & \multicolumn{1}{c|}{0.1774} & 0.3593 & 0.1976 & 0.6140 & 0.3473 \\
\multicolumn{1}{c|}{} & w/o shared-level & 0.3644 & \multicolumn{1}{c|}{0.1997} & 0.3971 & \multicolumn{1}{c|}{0.2268} & 0.3309 & \multicolumn{1}{c|}{0.1768} & 0.3574 & 0.1962 & 0.6121 & 0.3467 \\
\multicolumn{1}{c|}{} & w/o confounder & 0.3689 & \multicolumn{1}{c|}{0.2054} & 0.4032 & \multicolumn{1}{c|}{0.2277} & 0.3345 & \multicolumn{1}{c|}{0.1781} & 0.3618 & 0.1989 & 0.6166 & 0.3489 \\
\multicolumn{1}{c|}{} & w/ direct-LLM & 0.3816 & \multicolumn{1}{c|}{0.2169} & 0.4163 & \multicolumn{1}{c|}{0.2381} & 0.3462 & \multicolumn{1}{c|}{0.1843} & 0.3745 & 0.2091 & 0.6385 & 0.3624 \\
\multicolumn{1}{c|}{} & w/ gpt2glm & 0.3883 & \multicolumn{1}{c|}{0.2194} & 0.4240 & \multicolumn{1}{c|}{0.2425} & 0.3521 & \multicolumn{1}{c|}{0.1937} & 0.3807 & 0.2163 & 0.6488 & 0.3742 \\
\multicolumn{1}{c|}{} & w/ gpt2qwen & 0.3912 & \multicolumn{1}{c|}{0.2238} & 0.4266 & \multicolumn{1}{c|}{0.2453} & 0.3544 & \multicolumn{1}{c|}{0.1948} & 0.3851 & 0.2179 & 0.6536 & 0.3795 \\ \hline
\multicolumn{2}{c|}{\textbf{Improvement (CICDOR vs. best baselines)}} & 6.41\% & \multicolumn{1}{c|}{9.48\%} & 5.58\% & \multicolumn{1}{c|}{8.91\%} & 6.83\% & \multicolumn{1}{c|}{10.29\%} & 7.06\% & 10.09\% & 5.53\% & 8.32\% \\ \hline
\end{tabular}%
}
\vspace{-0.2in}
\end{table}

\item[(3)] While existing Causality-based OOD Recommendation methods consider knowledge that remains invariant across different distributions, they overlook the potential benefits of leveraging cross-domain invariant knowledge to enhance OOD recommendation in the target domain, thus yielding suboptimal results. Our CICDOR outperforms the best-performing Causality-based OOD Recommendation baseline, CausalDiffRec \cite{zhao2025graph}, by an average of 9.66\% and 13.15\% w.r.t. HR@10 and NDCG@10, respectively. The improvements stem from CICDOR's ability to utilize source domain data to learn the domain-shared causal structure, enabling the inference of domain-shared causal-invariant user preferences that can be transferred to facilitate OOD recommendation in the target domain;

\item[(4)] Among Single-target CDR baselines, CDCOR \cite{li2024cross} represents a best-performing single-level causal baseline specifically designed for OOD scenarios in CDR. Specifically, CDCOR only focuses on the domain-shared causal structure while neglecting domain-specific causal structures. Our CICDOR significantly outperforms CDCOR by an average of 10.34\% and 14.15\% w.r.t. HR@10 and NDCG@10, respectively. This improvement stems from CICDOR's dual-level causal design: while the domain-shared causal structure captures invariant knowledge for cross-domain transfer to mitigate both CDDS and SDDS, the domain-specific causal structure models invariant causal relationships within the target domain to further mitigate SDDS. The integration of both causal levels enables CICDOR to derive comprehensive causal-invariant preferences, leading to superior robustness across diverse OOD scenarios where both types of distribution shifts co-exist.

\item[(5)] Compared with the best-performing baseline, DT3OR \cite{yang2025dual}, our CICDOR achieves average improvements of 6.34\% and 9.75\% w.r.t. HR@10 and NDCG@10, respectively. These improvements stem from two key innovations. First, our dual-level causal preference learning module simultaneously captures both domain-shared and domain-specific causal-invariant user preferences, effectively addressing both CDDS and SDDS. This approach differs from existing single-target CDR baselines that either completely overlook SDDS (e.g., PTUPCDR \cite{zhu2022personalized} and CUT \cite{li2024aiming}) or only partially mitigate it by solely leveraging the domain-shared causal structure while neglecting the domain-specific causal structure in the target domain (e.g., CDCOR \cite{li2024cross}). Second, our LLM-guided confounder discovery module accurately identifies and extracts observed confounders from user reviews by leveraging LLM's knowledge and reasoning capabilities. By deconfounding these observed confounders, we ensure that the inferred causal-invariant user preferences can reflect users' true preferences rather than being influenced by confounding bias. The combination of these two complementary modules enables CICDOR to capture debiased comprehensive causal-invariant user preferences, resulting in superior cross-domain OOD recommendation performance across various OOD settings.
 \end{itemize}

\subsection{Ablation Study (for RQ2)}
\label{sec:V-C}
To evaluate the contribution of each component in improving the OOD recommendation performance of our model, we construct six variants of CICDOR and conduct the ablation study across various OOD settings (OOD \#2 and OOD \#3) on two datasets.

\subsubsection{\textbf{Impact of Dual-level Causal Preference Learning}}
To clearly disentangle the contributions of domain-specific and domain-shared causal structures in driving performance gains, we examine how each causal level addresses different types of distribution shifts: domain-specific causal structures primarily target SDDS, while domain-shared causal structures focus on CDDS. To empirically examine these roles, we construct three variants: \textbf{w/o dual-level} by removing the entire dual-level causal preference learning module, \textbf{w/o specific-level} by removing the domain-specific causal structure learning component, and \textbf{w/o shared-level} by removing the domain-shared causal structure learning component. From Table \ref{baseline_comparison}, we can observe that our CICDOR model outperforms \textbf{w/o dual-level} with an average improvement of 19.92\%. This significant performance difference demonstrates that the dual-level causal preference learning module is well suited for addressing distribution shifts in CDR. Without this module, the model lacks the ability to infer causal-invariant user preferences that remain invariant across different distributions. By learning the underlying causal structures and inferring causal-invariant user preferences, our CICDOR can achieve superior OOD recommendation performance in scenarios where both CDDS and SDDS co-exist.

Furthermore, the results show that our CICDOR outperforms \textbf{w/o specific-level} and \textbf{w/o shared-level} with average improvements of 9.32\% and 9.81\%, respectively. These comparable performance differences indicate that both levels of causal structure learning are essential and contribute almost equally to the model's overall effectiveness. By learning the domain-specific causal structure in the target domain, CICDOR infers domain-specific causal-invariant user preferences that can mitigate SDDS. Meanwhile, by learning the domain-shared causal structure, CICDOR infers domain-shared causal-invariant user preferences that facilitate invariant knowledge transfer from the source domain, thus effectively addressing CDDS and providing valuable support for OOD recommendation in the target domain. The fusion of both levels of causal-invariant user preferences yields comprehensive causal-invariant user preferences, thereby enabling our CICDOR to obtain better recommendation performance under various OOD settings. 

\subsubsection{\textbf{Impact of LLM-guided Confounder Discovery}}
We modify CICDOR to form a variant, namely \textbf{w/o confounder}, by removing the LLM-guided confounder discovery module. From Table \ref{baseline_comparison}, we can observe that our CICDOR model outperforms \textbf{w/o confounder} with an average improvement of 8.53\%. This demonstrates that extracting observed confounders for subsequent deconfounding is crucial for capturing accurate comprehensive causal-invariant user preferences, thereby achieving better OOD recommendation performance.

Moreover, we design another variant, namely \textbf{w/ direct-LLM}, by replacing our LLM-guided confounder discovery module with a simplified approach that directly uses gpt-4o-mini to extract confounders from user reviews in a single step. We find that the complete CICDOR model outperforms \textbf{w/ direct-LLM} by an average of 4.36\%. This demonstrates the effectiveness of our module over simply relying on LLM alone. Without the iterative process involving variable proposal, review annotation, variable refinement and causal feedback, simply using LLM to extract observed confounders may generate hallucinations or identify spurious variables as confounders. Our LLM-guided confounder discovery module addresses this limitation by providing theoretical guidance CI tests and the FCI algorithm, which help filter and refine the variables before extracting confounders. This process ensures that extracted confounders are causally valid and accurate, leading to more effective deconfounding and ultimately better OOD recommendation performance.

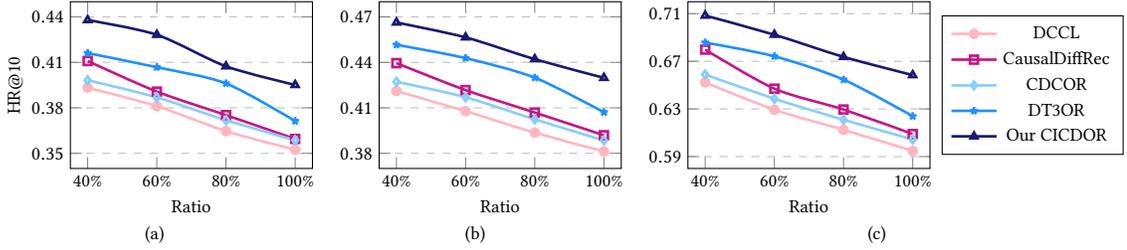
\begin{figure}[t]
 \setlength{\belowcaptionskip}{-0.12in}
 \centering
 \footnotesize
 \subfigure[]{
  \begin{tikzpicture}
  \begin{axis}[
  width=4.8cm,
  height=3.8cm,
  ylabel={HR@10},
  ylabel style ={yshift=-0.5cm},
  xlabel={Ratio},
  xlabel style ={yshift=0.2cm},
  xmin=35, xmax=105,
  ymin=0.34, ymax=0.45,
  xtick={40,60,80,100},
  xticklabels={40\%,60\%,80\%,100\%},
  yticklabel style={/pgf/number format/.cd,fixed,precision=3},
  ytick={0,0.35,0.38,0.41,0.44},
  scaled ticks=false,
  ymajorgrids=true,
  grid style=dashed,
  ]
  \addplot[color=red3,
  mark=*,
  mark options={solid},
  line width=1pt,mark size=1.5pt,
  smooth] coordinates {
   (40,0.3933)
   (60,0.3811)
   (80,0.3646)
   (100,0.3524)
   };
  \addplot[ color=red1,
  mark=square,
  mark options={solid},
  line width=1pt,mark size=1.5pt,
  smooth] coordinates {
   (40,0.4109)
   (60,0.3907)
   (80,0.3752)
   (100,0.3595)
   };
  \addplot[ color=blue3,
  mark=diamond,
  mark options={solid},
  line width=1pt,mark size=1.5pt,
  smooth] coordinates {
   (40,0.3982)
   (60,0.3869)
   (80,0.3718)
   (100,0.3586)
   };
  \addplot[ color=blue2,
  mark=star,
  mark options={solid},
  line width=1pt,mark size=1.5pt,
  smooth] coordinates {
   (40,0.4161)
   (60,0.4068)
   (80,0.3961)
   (100,0.3713)
   };
   \addplot[ color=blue1,
  mark=triangle,
  mark options={solid},
  line width=1pt,mark size=1.5pt,
  smooth] coordinates {
   (40,0.4379)
   (60,0.4282)
   (80,0.4074)
   (100,0.3951)
   };
  \end{axis}
  \end{tikzpicture}}
  \hspace{0in}
 \subfigure[]{
  \begin{tikzpicture}
  \begin{axis}[
  width=4.8cm,
  height=3.8cm,
  xlabel={Ratio},
  xlabel style ={yshift=0.2cm},
  xmin=35, xmax=105,
  ymin=0.37, ymax=0.48,
  xtick={40,60,80,100},
  xticklabels={40\%,60\%,80\%,100\%},
  yticklabel style={/pgf/number format/.cd,fixed,precision=3},
  ytick={0,0.38,0.41,0.44,0.47},
  scaled ticks=false,
  ymajorgrids=true,
  grid style=dashed,
  ]
  \addplot [color=red3,
  mark=*,
  mark options={solid},
  line width=1pt,mark size=1.5pt,
  smooth]coordinates {
   (40,0.4211)
   (60,0.4078)
   (80,0.3936)
   (100,0.3812)
   };
  \addplot [color=red1,
  mark=square,
  mark options={solid},
  line width=1pt,mark size=1.5pt,
  smooth]coordinates {
   (40,0.4395)
   (60,0.4217)
   (80,0.4069)
   (100,0.3919)
   };
  \addplot [color=blue3,
  mark=diamond,
  mark options={solid},
  line width=1pt,mark size=1.5pt,
  smooth]coordinates {
   (40,0.4272)
   (60,0.4170)
   (80,0.4024)
   (100,0.3885)
   };
  \addplot [ color=blue2,
  mark=star,
  mark options={solid},
  line width=1pt,mark size=1.5pt,
  smooth]coordinates {
   (40,0.4516)
   (60,0.4428)
   (80,0.4299)
   (100,0.4071)
   };
   \addplot[ color=blue1,
  mark=triangle,
  mark options={solid},
  line width=1pt,mark size=1.5pt,
  smooth] coordinates {
   (40,0.4663)
   (60,0.4565)
   (80,0.4421)
   (100,0.4298)
   };
  \end{axis}
  \end{tikzpicture}}
  \hspace{0in}
 \subfigure[]{
  \begin{tikzpicture}
  \begin{axis}[
  width=4.8cm,
  height=3.8cm,
  xlabel={Ratio},
  xlabel style ={yshift=0.2cm},
  xmin=35, xmax=105,
  ymin=0.58, ymax=0.72,
  xtick={40,60,80,100},
  xticklabels={40\%,60\%,80\%,100\%},
  yticklabel style={/pgf/number format/.cd,fixed,precision=3},
  ytick={0,0.59,0.63,0.67,0.71},
  scaled ticks=false,
  legend style={at={(1.8,0.5)}, anchor=east,legend columns=1, draw=black},
  ymajorgrids=true,
  grid style=dashed,
  ]
  \addplot [ color=red3,
  mark=*,
  mark options={solid},
  line width=1pt,mark size=1.5pt,
  smooth]coordinates {
   (40,0.6523)
   (60,0.6293)
   (80,0.6124)
   (100,0.5948)
   };
  \addplot [ color=red1,
  mark=square,
  mark options={solid},
  line width=1pt,mark size=1.5pt,
  smooth]coordinates {
   (40,0.6798)
   (60,0.6470)
   (80,0.6295)
   (100,0.6089)
   };
  \addplot [ color=blue3,
  mark=diamond,
  mark options={solid},
  line width=1pt,mark size=1.5pt,
  smooth]coordinates {
   (40,0.6589)
   (60,0.6385)
   (80,0.6209)
   (100,0.6044)
   };
  \addplot [ color=blue2,
  mark=star,
  mark options={solid},
  line width=1pt,mark size=1.5pt,
  smooth]coordinates {
   (40,0.6857)
   (60,0.6742)
   (80,0.6546)
   (100,0.6239)
   };
   \addplot[ color=blue1,
  mark=triangle,
  mark options={solid},
  line width=1pt,mark size=1.5pt,
  smooth] coordinates {
   (40,0.7084)
   (60,0.6923)
   (80,0.6737)
   (100,0.6584)
   };
   \legend{DCCL, CausalDiffRec, CDCOR, DT3OR, Our CICDOR}
  \end{axis}
  \end{tikzpicture}}
  \vspace{-0.2in}
 \caption{Performance comparison (HR@10) between CICDOR and baselines under different degrees of distribution shift: (a) Douban-Movie $\rightarrow$ Douban-Book OOD \#2, (b) Douban-Movie $\rightarrow$ Douban-Book OOD \#3, and (c) Amazon-Elec $\rightarrow$ Amazon-Cloth OOD \#2.}
 \label{fig:different_degrees}
 \vspace{-0.1in}
\end{figure}

To evaluate the robustness across different LLM backends, we introduce two additional variants, denoted as \textbf{w/ gpt2glm} and \textbf{w/ gpt2qwen}, where the LLM backend in the confounder discovery module is replaced from gpt-4o-mini to glm-4-9b-chat\footnote{https://huggingface.co/THUDM/glm-4-9b-chat} and Qwen3-8b\footnote{https://huggingface.co/Qwen/Qwen3-8B}, respectively. We can observe that compared to our CICDOR, \textbf{w/ gpt2glm} and \textbf{w/ gpt2qwen} exhibit average performance drops of 2.05\% and 0.98\%, respectively. These results show that while stronger LLMs enable more accurate confounder extraction, our CICDOR is not overly dependent on specific LLM capabilities. In our implementation, we select gpt-4o-mini to balance performance benefits with economic considerations, enabling us to extract accurate confounders for effective deconfounding and better OOD recommendation.

\subsection{Stability Analysis (for RQ3)}
\label{sec:V-D}
\subsubsection{\textbf{Impact of varying degrees of distribution shift}}
To investigate our model's recommendation performance under varying degrees of distribution shift, we compared CICDOR with the best-performing baseline from each group across different shift intensities. For user degree shift (OOD \#2), we control the proportion of samples from users with high interaction degrees in the test set, ranging from 40\% to 100\%, with the remaining portions filled by samples from users with low interaction degrees. Similarly, for region shift (OOD \#3), we control the proportion of samples from Beijing users in the test set from 40\% to 100\%, with non-Beijing user samples filling the remainder.

As shown in Fig. \ref{fig:different_degrees}, the recommendation performance\footnote{Due to space limitations, Fig. \ref{fig:different_degrees} and Fig. \ref{different_parameter} only show results for Douban-Movie $\rightarrow$ Douban-Book and Amazon-Elec $\rightarrow$ Amazon-Cloth. Similar trends are also observed in the Douban-Movie $\rightarrow$ Douban-Music experiments.} of all models decreases as distribution shift intensifies. However, CICDOR consistently achieves the best performance across all shift ratios. Moreover, CICDOR exhibits the most gradual performance decline as distribution shift increases, demonstrating superior stability in OOD generalization. Both the superior performance and enhanced stability can be attributed to CICDOR's ability to learn dual-level invariant causal structures from cross-domain knowledge, extract accurate observed confounders, and mitigate these confounders' negative effects on the inferred causal-invariant user preferences. This enables our CICDOR to produce more accurate and robust recommendations across various OOD settings.

\subsubsection{\textbf{Impact of different overlapping user ratios}}
To further investigate the stability of our proposed CICDOR under partial user overlap, we conduct additional experiments in the region shift setting (OOD \#3). Specifically, we vary the overlapping user ratio from 100\% to 30\% and compare our proposed CICDOR with CDCOR, the best-performing single-target CDR baseline. For each ratio, the corresponding proportion of originally overlapping users retain interaction data in both domains during training, while the remaining users retain only their interaction data in the target domain. This design allows us to examine the impact of reduced overlapping users without introducing additional sparsity in the target domain.

\begin{figure}[t]
 \setlength{\belowcaptionskip}{-0.12in}
 \centering
 \footnotesize
 \subfigure[]{
  \begin{tikzpicture}
  \begin{axis}[
  width=6.0cm,
  height=4.6cm,
  ylabel={HR@10},
  ylabel style ={yshift=-0.5cm},
  xlabel={Overlapping User Ratio},
  xlabel style ={yshift=0.2cm},
  xmin=25, xmax=105,
  ymin=0.285, ymax=0.445,
  xtick={30,40,50,60,70,80,90,100},
  xticklabels={100\%,90\%,80\%,70\%,60\%,50\%,40\%,30\%},
  yticklabel style={/pgf/number format/.cd,fixed,precision=3},
  ytick={0,0.29,0.34,0.39,0.44},
  scaled ticks=false,
  legend style={at={(0.45,0.18)}, anchor=east,legend columns=1, draw=black},
  ymajorgrids=true,
  grid style=dashed,
  ]
  \addplot[color=red3,
  mark=*,
  mark options={solid},
  line width=1pt,mark size=1.5pt,
  smooth] coordinates {
   (30,0.3885)
   (40,0.3819)
   (50,0.3710)
   (60,0.3638)
   (70,0.3545)
   (80,0.3409)
   (90,0.3231)
   (100,0.3043)
   };
  \addplot[ color=red1,
  mark=square,
  mark options={solid},
  line width=1pt,mark size=1.5pt,
  smooth] coordinates {
   (30,0.4298)
   (40,0.4264)
   (50,0.4213)
   (60,0.4132)
   (70,0.4033)
   (80,0.3943)
   (90,0.3815)
   (100,0.3664)
   };
   \legend{CDCOR, CICDOR}
  \end{axis}
  \end{tikzpicture}}
  \hspace{0in}
 \subfigure[]{
  \begin{tikzpicture}
  \begin{axis}[
  width=6.0cm,
  height=4.6cm,
  ylabel={NDCG@10},
  ylabel style ={yshift=-0.5cm},
  xlabel={Overlapping User Ratio},
  xlabel style ={yshift=0.2cm},
  xmin=25, xmax=105,
  ymin=0.165, ymax=0.265,
  xtick={30,40,50,60,70,80,90,100},
  xticklabels={100\%,90\%,80\%,70\%,60\%,50\%,40\%,30\%},
  yticklabel style={/pgf/number format/.cd,fixed,precision=3},
  ytick={0,0.17,0.20,0.23,0.26},
  scaled ticks=false,
  legend style={at={(0.45,0.18)}, anchor=east,legend columns=1, draw=black},
  ymajorgrids=true,
  grid style=dashed,
  ]
  \addplot [ color=red3,
  mark=*,
  mark options={solid},
  line width=1pt,mark size=1.5pt,
  smooth]coordinates {
   (30,0.2196)
   (40,0.2169)
   (50,0.2066)
   (60,0.1995)
   (70,0.1956)
   (80,0.1828)
   (90,0.1749)
   (100,0.1727)
   };
  \addplot [ color=red1,
  mark=square,
  mark options={solid},
  line width=1pt,mark size=1.5pt,
  smooth]coordinates {
   (30,0.2493)
   (40,0.2452)
   (50,0.2411)
   (60,0.2367)
   (70,0.2278)
   (80,0.2259)
   (90,0.2167)
   (100,0.2001)
   };
   \legend{CDCOR, CICDOR}
  \end{axis}
  \end{tikzpicture}}
  \hspace{0in}
 \subfigure[]{
  \begin{tikzpicture}
  \begin{axis}[
  width=6.0cm,
  height=4.6cm,
  ylabel={HR@10},
  ylabel style ={yshift=-0.5cm},
  xlabel={Overlapping User Ratio},
  xlabel style ={yshift=0.2cm},
  xmin=25, xmax=105,
  ymin=0.245, ymax=0.405,
  xtick={30,40,50,60,70,80,90,100},
  xticklabels={100\%,90\%,80\%,70\%,60\%,50\%,40\%,30\%},
  yticklabel style={/pgf/number format/.cd,fixed,precision=3},
  ytick={0,0.25,0.30,0.35,0.40},
  scaled ticks=false,
  legend style={at={(0.45,0.18)}, anchor=east,legend columns=1, draw=black},
  ymajorgrids=true,
  grid style=dashed,
  ]
  \addplot [ color=red3,
  mark=*,
  mark options={solid},
  line width=1pt,mark size=1.5pt,
  smooth]coordinates {
   (30,0.3543)
   (40,0.3457)
   (50,0.3342)
   (60,0.3226)
   (70,0.3077)
   (80,0.2955)
   (90,0.2732)
   (100,0.2607)
   };
  \addplot [ color=red1,
  mark=square,
  mark options={solid},
  line width=1pt,mark size=1.5pt,
  smooth]coordinates {
   (30,0.3882)
   (40,0.3822)
   (50,0.3739)
   (60,0.3653)
   (70,0.3543)
   (80,0.3451)
   (90,0.3309)
   (100,0.3171)
   };
   \legend{CDCOR, CICDOR}
  \end{axis}
  \end{tikzpicture}}
  \hspace{0in}
 \subfigure[]{
  \begin{tikzpicture}
  \begin{axis}[
  width=6.0cm,
  height=4.6cm,
  ylabel={NDCG@10},
  ylabel style ={yshift=-0.5cm},
  xlabel={Overlapping User Ratio},
  xlabel style ={yshift=0.2cm},
  xmin=25, xmax=105,
  ymin=0.135, ymax=0.235,
  xtick={30,40,50,60,70,80,90,100},
  xticklabels={100\%,90\%,80\%,70\%,60\%,50\%,40\%,30\%},
  yticklabel style={/pgf/number format/.cd,fixed,precision=3},
  ytick={0,0.14,0.17,0.20,0.23},
  scaled ticks=false,
  legend style={at={(0.45,0.18)}, anchor=east,legend columns=1, draw=black},
  ymajorgrids=true,
  grid style=dashed,
  ]
  \addplot [ color=red3,
  mark=*,
  mark options={solid},
  line width=1pt,mark size=1.5pt,
  smooth]coordinates {
   (30,0.1956)
   (40,0.1844)
   (50,0.1782)
   (60,0.1749)
   (70,0.1701)
   (80,0.1636)
   (90,0.1478)
   (100,0.1419)
   };
  \addplot [ color=red1,
  mark=square,
  mark options={solid},
  line width=1pt,mark size=1.5pt,
  smooth]coordinates {
   (30,0.2194)
   (40,0.2171)
   (50,0.2088)
   (60,0.1999)
   (70,0.1955)
   (80,0.1843)
   (90,0.1766)
   (100,0.1722)
   };
   \legend{CDCOR, CICDOR}
  \end{axis}
  \end{tikzpicture}}
  \vspace{-0.2in}
 \caption{Performance comparison between our proposed CICDOR and the best-performing single-target CDR baseline CDCOR under different overlapping user ratios: (a)-(b): Douban-Movie $\rightarrow$ Douban-Book OOD \#3. (c)-(d) Douban-Movie $\rightarrow$ Douban-Music OOD \#3.}
 \label{fig:different_ratios}
 \vspace{-0.1in}
\end{figure}

From Fig. \ref{fig:different_ratios}, we have the following observations. (1) As the overlapping user ratio decreases, the recommendation performance of both our proposed CICDOR and CDCOR gradually declines, indicating that overlapping users play an important role in effective cross-domain preference transfer. (2) Nevertheless, our proposed CICDOR consistently outperforms CDCOR across all overlapping user ratios, indicating its superior recommendation performance under the same degree of user overlap. More importantly, our proposed CICDOR is less sensitive than CDCOR to the reduction of the overlapping user ratio. Specifically, in terms of HR@10, when the overlapping user ratio decreases to 30\%, the performance of our proposed CICDOR drops by 16.53\% compared to its performance under 100\% overlap, whereas that of CDCOR drops by 24.05\%. This smaller relative degradation further suggests that our proposed CICDOR is more robust to reduced user overlap. This is because, although both methods utilize domain-shared causal structure learning for cross-domain preference transfer, our proposed CICDOR additionally models domain-specific causal-invariant user preferences within the target domain. When the overlapping user ratio is reduced, the domain-shared preference transfer is weakened, while the domain-specific causal-invariant user preferences provide complementary support for OOD recommendation, thereby alleviating the performance degradation under partial user overlap. (3) Notably, across all four subfigures in Figs. \ref{fig:different_ratios}(a)-\ref{fig:different_ratios}(d), our proposed CICDOR with 60\% overlapping users achieves performance comparable to or even superior to CDCOR with full user overlap (100\%). This further suggests that our proposed CICDOR is not restricted to the full user overlap scenario and can still achieve competitive OOD recommendation performance under reduced overlapping user ratios.

\subsection{Parameter Sensitivity (for RQ4)}
\label{sec:V-E}
\subsubsection{\textbf{Impact of the number of cluster centroids $J$}}
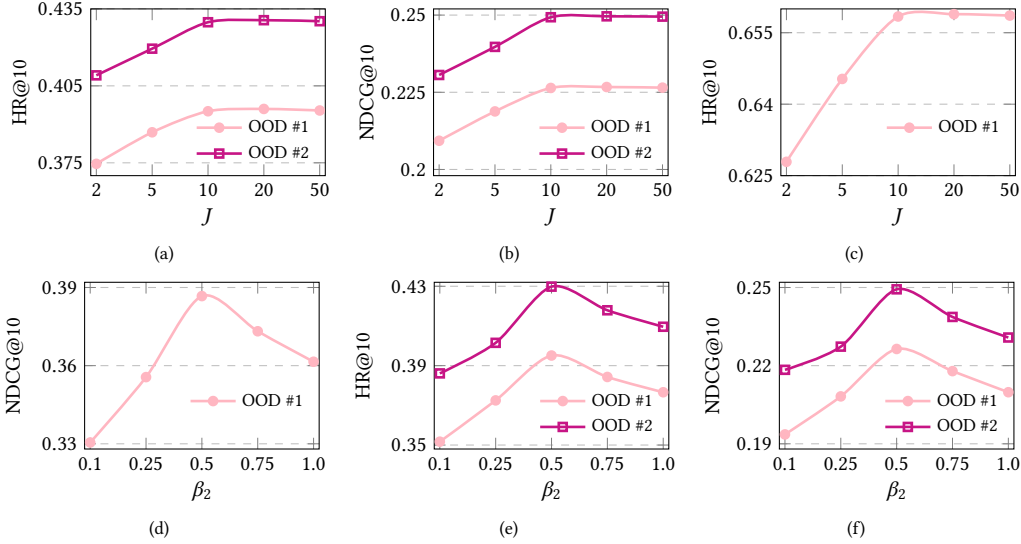
\begin{figure}[t]
 \setlength{\belowcaptionskip}{-0.12in}
 \centering
 \small
  \subfigure[]{
  \begin{tikzpicture}
  \begin{axis}[
  width=4.7cm,
  height=3.8cm,
  xmin=0.9, xmax=5.1,
  ymin=0.370, ymax=0.435,
  ylabel={HR@10},
  ylabel style={yshift=-0.3cm, font=\small},
  xlabel={$J$},
  xlabel style={yshift=0.2cm},
  xtick={1,2,3,4,5},
  xticklabels={2,5,10,20,50},
  xticklabel style={font=\small},
  yticklabel style={font=\small, /pgf/number format/.cd, fixed, precision=3},
  ytick={0,0.375,0.405,0.435},
  scaled ticks=false,
  legend style={at={(0.70,0.40)}, font=\footnotesize, anchor=north,legend columns=1, draw=none, fill=none,},
  ymajorgrids=true,
  grid style=dashed,
  ]
  \addplot[color=red3,
  mark=*,
  mark options={solid},
  line width=1pt,mark size=1.5pt,
  smooth] coordinates {
   (1,0.3746)
   (2,0.3869)
   (3,0.3951)
   (4,0.3960)
   (5,0.3954)
   };
  \addplot[ color=red1,
  mark=square,
  mark options={solid},
  line width=1pt,mark size=1.5pt,
  smooth] coordinates {
   (1,0.4091)
   (2,0.4195)
   (3,0.4298)
   (4,0.4306)
   (5,0.4302)
   };
  \legend{OOD \#2, OOD \#3}
  \end{axis}
  \end{tikzpicture}}
  \hspace{0in}
  \vspace{-0.1in}
 \subfigure[]{
  \begin{tikzpicture}
  \begin{axis}[
  width=4.7cm,
  height=3.8cm,
  xmin=0.9, xmax=5.1,
  ymin=0.198, ymax=0.252,
  ylabel={NDCG@10},
  ylabel style={yshift=-0.36cm, font=\small},
  xlabel={$J$},
  xlabel style={yshift=0.2cm},
  xtick={1,2,3,4,5},
  xticklabels={2,5,10,20,50},
  xticklabel style={font=\small},
  yticklabel style={font=\small, /pgf/number format/.cd, fixed, precision=3},
  ytick={0,0.20,0.225,0.25},
  scaled ticks=false,
  legend style={at={(0.70,0.40)}, font=\footnotesize, anchor=north,legend columns=1, draw=none, fill=none,},
  ymajorgrids=true,
  grid style=dashed,
  ]
  \addplot [color=red3,
  mark=*,
  mark options={solid},
  line width=1pt,mark size=1.5pt,
  smooth]coordinates {
   (1,0.2093)
   (2,0.2188)
   (3,0.2264)
   (4,0.2267)
   (5,0.2265)
  };
  \addplot [color=red1,
  mark=square,
  mark options={solid},
  line width=1pt,mark size=1.5pt,
  smooth]coordinates {
   (1,0.2306)
   (2,0.2397)
   (3,0.2493)
   (4,0.2496)
   (5,0.2495)
  };
  \legend{OOD \#2, OOD \#3}
  \end{axis}
  \end{tikzpicture}}
  \hspace{0in}
  \subfigure[]{
  \begin{tikzpicture}
  \begin{axis}[
  width=4.7cm,
  height=3.8cm,
  xmin=0.9, xmax=5.1,
  ymin=0.625, ymax=0.660,
  ylabel={HR@10},
  ylabel style={yshift=-0.3cm, font=\small},
  xlabel={$J$},
  xlabel style={yshift=0.2cm},
  xtick={1,2,3,4,5},
  xticklabels={2, 5, 10, 20, 50},
  xticklabel style={font=\small},
  yticklabel style={font=\small, /pgf/number format/.cd, fixed, precision=3},
  ytick={0,0.625,0.640,0.655},
  scaled ticks=false,
  legend style={at={(0.70,0.40)}, font=\footnotesize, anchor=north,legend columns=1, draw=none, fill=none,},
  ymajorgrids=true,
  grid style=dashed,
  ]
  \addplot[color=red3,
  mark=*,
  mark options={solid},
  line width=1pt,mark size=1.5pt,
  smooth] coordinates {
   (1,0.6279)
   (2,0.6453)
   (3,0.6584)
   (4,0.6589)
   (5,0.6586)
   };
  \legend{OOD \#2}
  \end{axis}
  \end{tikzpicture}}
  \hspace{0in}
  \subfigure[]{
  \begin{tikzpicture}
  \begin{axis}[
  width=4.7cm,
  height=3.8cm,
  xmin=0.9, xmax=5.1,
  ymin=0.328, ymax=0.392,
  ylabel={NDCG@10},
  ylabel style={yshift=-0.3cm, font=\small},
  xlabel={$\beta_2$},
  xlabel style={yshift=0.2cm},
  xtick={1,2,3,4,5},
  xticklabels={0.1, 0.25, 0.5, 0.75, 1.0},
  xticklabel style={font=\small},
  yticklabel style={font=\small, /pgf/number format/.cd, fixed, precision=3},
  ytick={0,0.33,0.36,0.39},
  scaled ticks=false,
  legend style={at={(0.70,0.40)}, font=\footnotesize, anchor=north,legend columns=1, draw=none, fill=none,},
  ymajorgrids=true,
  grid style=dashed,
  ]
  \addplot [color=red3,
  mark=*,
  mark options={solid},
  line width=1pt,mark size=1.5pt,
  smooth]coordinates {
   (1,0.3305)
   (2,0.3556)
   (3,0.3867)
   (4,0.3732)
   (5,0.3615)
  };
  \legend{OOD \#2}
  \end{axis}
  \end{tikzpicture}}
  \hspace{0in}
 \subfigure[]{
  \begin{tikzpicture}
  \begin{axis}[
  width=4.7cm,
  height=3.8cm,
  xmin=0.9, xmax=5.1,
  ymin=0.348, ymax=0.432,
  ylabel={HR@10},
  ylabel style={yshift=-0.3cm, font=\small},
  xlabel={$\beta_2$},
  xlabel style={yshift=0.2cm},
  xtick={1,2,3,4,5},
  xticklabels={0.1, 0.25, 0.5, 0.75, 1.0},
  xticklabel style={font=\small},
  yticklabel style={font=\small, /pgf/number format/.cd, fixed, precision=3},
  ytick={0,0.35,0.39,0.43},
  scaled ticks=false,
  legend style={at={(0.70,0.40)}, font=\footnotesize, anchor=north,legend columns=1, draw=none, fill=none,},
  ymajorgrids=true,
  grid style=dashed,
  ]
  \addplot [color=red3,
  mark=*,
  mark options={solid},
  line width=1pt,mark size=1.5pt,
  smooth]coordinates {
   (1,0.3517)
   (2,0.3725)
   (3,0.3951)
   (4,0.3843)
   (5,0.3766)
  };
  \addplot [color=red1,
  mark=square,
  mark options={solid},
  line width=1pt,mark size=1.5pt,
  smooth]coordinates {
   (1,0.3861)
   (2,0.4015)
   (3,0.4298)
   (4,0.4179)
   (5,0.4096)
  };
  \legend{OOD \#2, OOD \#3}
  \end{axis}
  \end{tikzpicture}}
  \hspace{0in}
  \subfigure[]{
  \begin{tikzpicture}
  \begin{axis}[
  width=4.7cm,
  height=3.8cm,
  xmin=0.9, xmax=5.1,
  ymin=0.188, ymax=0.252,
  ylabel={NDCG@10},
  ylabel style={yshift=-0.36cm, font=\small},
  xlabel={$\beta_2$},
  xlabel style={yshift=0.2cm},
  xtick={1,2,3,4,5},
  xticklabels={0.1, 0.25, 0.5, 0.75, 1.0},
  xticklabel style={font=\small},
  yticklabel style={font=\small, /pgf/number format/.cd, fixed, precision=3},
  ytick={0,0.19,0.22,0.25},
  scaled ticks=false,
  legend style={at={(0.70,0.40)}, font=\footnotesize, anchor=north,legend columns=1, draw=none, fill=none,},
  ymajorgrids=true,
  grid style=dashed,
  ]
  \addplot[color=red3,
  mark=*,
  mark options={solid},
  line width=1pt,mark size=1.5pt,
  smooth] coordinates {
   (1,0.1936)
   (2,0.2082)
   (3,0.2264)
   (4,0.2179)
   (5,0.2098)
   };
  \addplot[ color=red1,
  mark=square,
  mark options={solid},
  line width=1pt,mark size=1.5pt,
  smooth] coordinates {
   (1,0.2184)
   (2,0.2273)
   (3,0.2493)
   (4,0.2387)
   (5,0.2308)
   };
  \legend{OOD \#2, OOD \#3}
  \end{axis}
  \end{tikzpicture}}
 \vspace{-0.2in}
 \caption{(a)-(b): Impact of the number of cluster centroids $J$ on Douban-Movie $\rightarrow$ Douban-Book. (c)-(d): Impact of $J$ and the weight of dual-level causal loss $\beta_2$ on Amazon-Elec $\rightarrow$ Amazon-Cloth. (e)-(f): Impact of $\beta_2$ on Douban-Movie $\rightarrow$ Douban-Book.}
 \label{different_parameter}
\vspace{-0.1in}
\end{figure}

We examine how varying the number of cluster centroids $J$ affects CICDOR's recommendation performance by testing $J=J^s=J^t$ across \{2, 5, 10, 20, 50\}. Figs. \ref{different_parameter}(a)-\ref{different_parameter}(c) present the results across different OOD settings (OOD \#2 and OOD \#3) on both Douban and Amazon datasets. We can observe that the recommendation performance of our CICDOR improves consistently as $J$ increases up to 10, confirming the effectiveness of confounders extracted through our LLM-guided confounder discovery module. At $J=10$, the observed confounders represented by these cluster centroids provide sufficient information for accurate deconfounding. Beyond this threshold, performance gains become negligible or even slightly decrease in some cases, suggesting that additional centroids may introduce noise rather than capture meaningful confounders. Although our confounder selection function has the ability to select relevant confounders, an excessive number of potential confounders appears counterproductive. Balancing model complexity and performance, we set $J=J^s=J^t=10$ for all experiments. This threshold indicates that approximately 10 key observed confounders exist in each domain, and our CICDOR can successfully extract them and then mitigate their negative effects on the inferred causal-invariant user preferences to improve the OOD recommendation performance.

\subsubsection{\textbf{Impact of the weight of dual-level causal loss $\beta_2$}}
\label{sec:5.5.2}
We investigate how the weight of dual-level causal loss $\beta_2$ affects our model's OOD recommendation performance by varying it across $\{0.1, 0.25, 0.5, 0.75, 1.0\}$. Figs. \ref{different_parameter}(d)-(f) illustrate that CICDOR achieves optimal OOD recommendation performance when $\beta_2=0.5$. This finding suggests that this specific value creates an ideal balance between the dual-level causal loss and recommendation loss. At this balanced point, our CICDOR can effectively learn domain-shared and domain-specific causal-invariant user preferences while maintaining recommendation accuracy. This balance enables our model to leverage invariant knowledge from the source domain to enhance OOD recommendation performance in the target domain, particularly when both CDDS and SDDS exist. When $\beta_2$ is either too large or too small, the model struggles to balance recommendation accuracy with OOD generalization capability, leading to suboptimal results. Based on these observations, we select $\beta_2=0.5$ for all our experiments to better optimize the recommendation losses and dual-level causal loss.

\subsection{Robustness Analysis (for RQ5)}
\label{sec:V-F}
To investigate CICDOR's robustness, we evaluate CICDOR under two types of perturbations in the LLM-guided confounder discovery module: annotation noise and prompt perturbations. For annotation noise analysis, we simulate imperfect labeling conditions by randomly corrupting 10\% and 20\% of the annotation values. For prompt sensitivity evaluation, we construct two variants: \textbf{Paraphrased} prompts reformulate task descriptions using semantically equivalent but structurally different instructions, while \textbf{Simplified} prompts reduce descriptive complexity by removing auxiliary explanations. Both variants preserve the core reasoning objectives.

\begin{table}[t]
\caption{Robustness analysis of CICDOR under annotation noise and prompt perturbations on the Amazon dataset (Elec $\rightarrow$ Cloth, OOD \#2).}
\label{tab:robustness_analysis}
\vspace{-0.1in}
\begin{tabular}{c|cccc}
\hline
Analysis Type                       & Condition   & HR@10           & NDCG@10         & Avg. Drop \\ \hline
\multirow{3}{*}{Annotation Noise}   & Original    & \textbf{0.6584} & \textbf{0.3867} & -         \\
                                    & 10\%        & 0.6514          & 0.3788          & 1.55\%    \\
                                    & 20\%        & 0.6426          & 0.3703          & 1.80\%    \\ \hline
\multirow{3}{*}{Prompt Sensitivity} & Original    & \textbf{0.6584} & \textbf{0.3867} & -         \\
                                    & Paraphrased & 0.6547          & 0.3812          & 0.99\%    \\
                                    & Simplified  & 0.6421          & 0.3735          & 1.97\%    \\ \hline
\end{tabular}
\vspace{-0.2in}
\end{table}

Table \ref{tab:robustness_analysis} reports the robustness analysis results on Amazon-Elec $\rightarrow$ Amazon-Cloth task under the OOD \#2 setting. For annotation noise analysis, CICDOR exhibits limited performance degradation, with average drops of 1.55\% and 1.80\% when injecting 10\% and 20\% noise, respectively. These results indicate that the overall performance exhibits low sensitivity to annotation noise introduced during variable annotation. For prompt sensitivity analysis, CICDOR maintains stable performance across different prompt variants. The \textbf{Paraphrased} prompts result in less than 1\% performance degradation, while the \textbf{Simplified} prompts lead to a slightly larger drop of around 2\%. Importantly, even when prompt perturbations are consistently applied across multiple LLM-invoked steps, the resulting performance degradation remains modest, demonstrating the robustness of our CICDOR to perturbations in the LLM-guided confounder discovery module.

\subsection{Case Study (for RQ6)}
\label{sec:V-G}
\begin{figure*}[t]
\centering
\begin{minipage}{\linewidth}
  \centering
  \includegraphics[width=\textwidth]{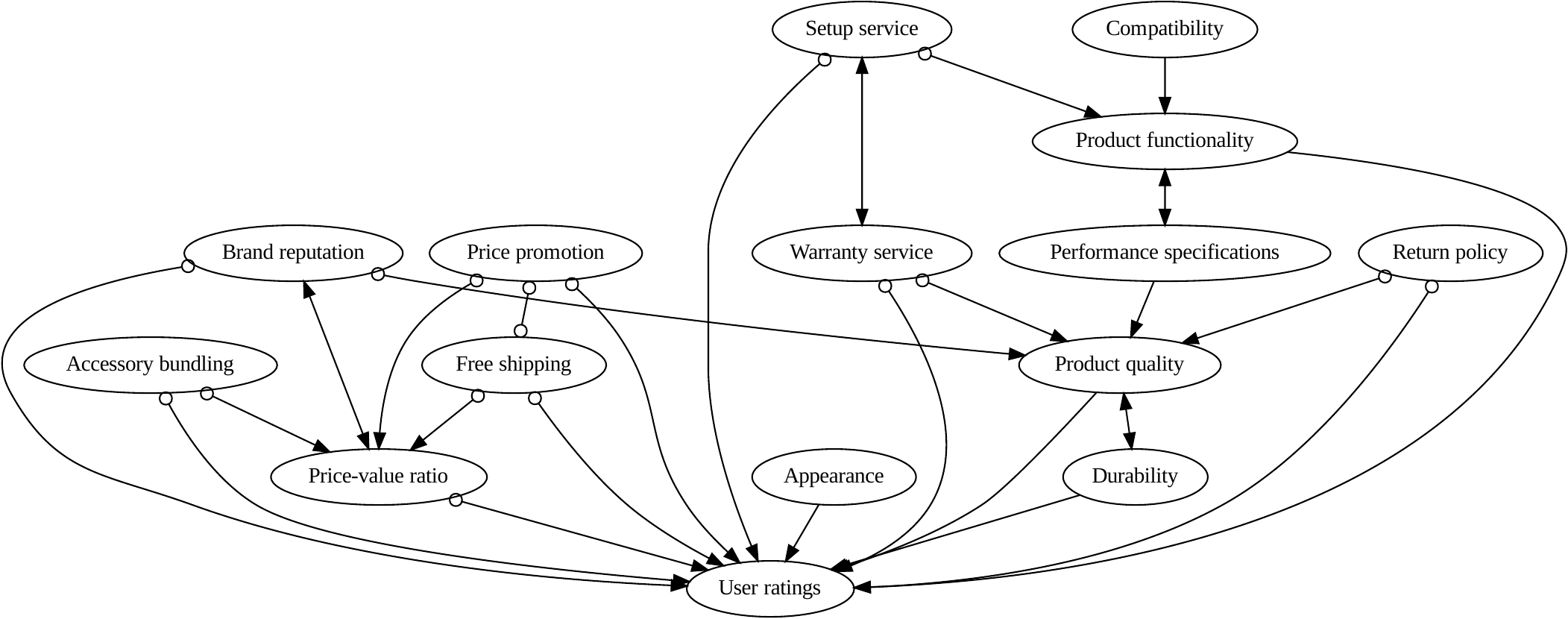}
  \centerline{(a) Electronics domain}\medskip
\end{minipage}

\vspace{0.2cm}

\begin{minipage}{\linewidth}
  \centering
  \includegraphics[width=0.91\textwidth]{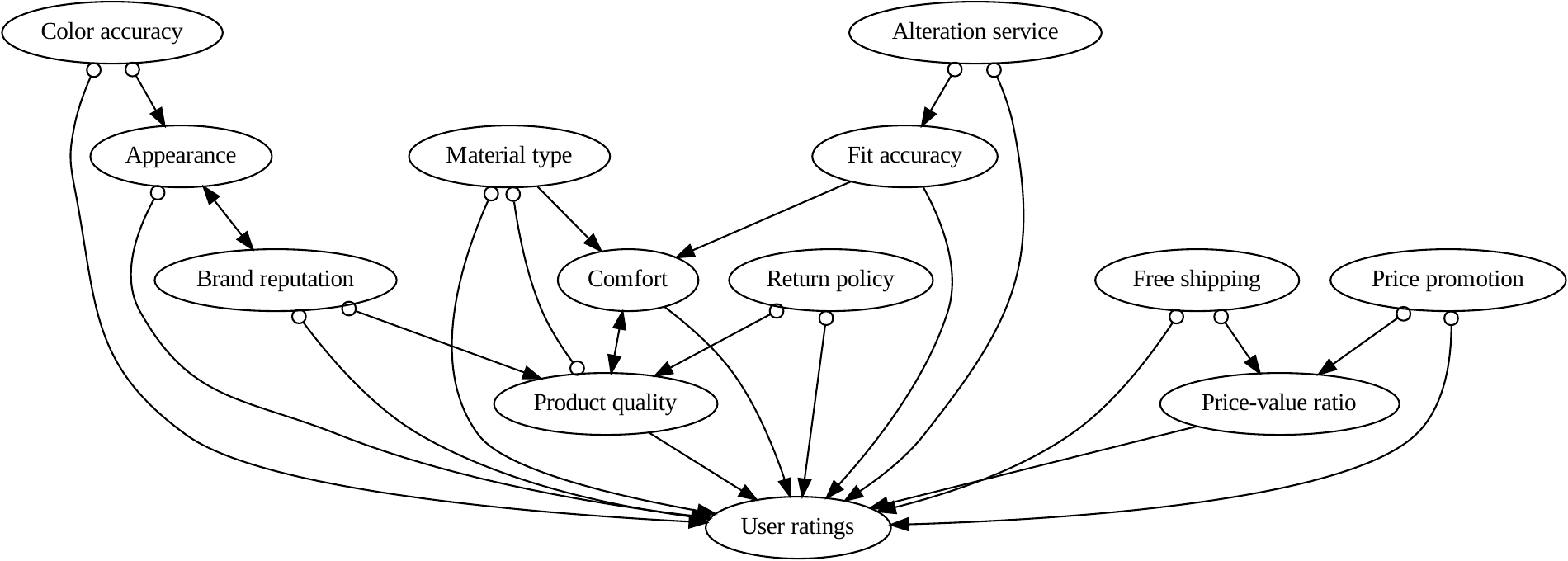}
  \centerline{(b) Clothing domain}\medskip
\end{minipage}
\vspace{-0.1in}
\caption{Learned causal structures discovered by CICDOR in two Amazon domains.}
\vspace{-0.2in}
\label{fig:causal_structures}
\end{figure*}

To understand the causal mechanisms underlying user-item interactions, we apply our CICDOR framework to two representative domains from the Amazon review dataset: Electronics and Clothing. As part of CICDOR, the LLM-guided confounder discovery module extracts high-level causal variables from user reviews and learns their causal relationships via the FCI algorithm. The learned causal structures are visualized in Fig. \ref{fig:causal_structures}. In both domains, user ratings serve as a proxy for the target variable, indicating whether a user interacts with an item.

Taking the Electronics domain as an example, the learned causal structure reveals several key causal relationships. As illustrated in Fig. \ref{fig:causal_structures}(a), core causal variables such as \emph{Appearance}, \emph{Price-value ratio}, \emph{Product quality}, and \emph{Durability} directly influence user ratings. Additionally, upstream variables like \emph{Performance specifications} and \emph{Product functionality} indirectly affect user ratings through their impact on perceived product quality, forming clear causal pathways.

In addition to these direct and indirect causal effects, the learned structure also reveals a set of causal variables that simultaneously influence user preferences and user-item interactions. Such variables, referred to as observed confounders, can be broadly categorized into two types: single-domain confounders (SDCs), which affect user behavior within a specific domain, and cross-domain confounders (CDCs), which affect user behavior across both source and target domains in CDR.

In the Electronics domain, several SDCs are identified, including \emph{Setup service}, \emph{Warranty service}, and \emph{Accessory bundling}. For example, \emph{Setup service} affects user ratings both directly and indirectly through its impact on \emph{Product functionality} (Fig. \ref{fig:causal_structures}(a)). While users' purchasing decisions are primarily driven by their true preferences, a confounder such as \emph{Setup service} serves as a catalyst that influences their interaction decisions. This type of confounder is unique to the Electronics domain and does not appear in the Clothing domain, making it a representative SDC.

By contrast, CDCs exist in both domains. For instance, \emph{Price promotion} is a CDC, as it affects user ratings both directly and indirectly through its impact on \emph{Price-value ratio} in both the Electronics (Fig. \ref{fig:causal_structures}(a)) and Clothing (Fig. \ref{fig:causal_structures}(b)) domains. Similar to \emph{Setup service}, users do not make purchasing decisions solely due to price promotions. Their true preferences remain the primary driver, while such confounders act as catalysts that influence user behavior in both domains. Other CDCs include \emph{Free shipping} and \emph{Return policy}, which consistently appear in the causal structures of both domains.

Overall, the above case study demonstrates that the causal structures learned by CICDOR capture meaningful and intuitive relationships among causal variables. Both direct and indirect causal pathways align well with general user decision-making processes. Moreover, observed confounders, including both SDCs and CDCs, are successfully identified as candidate causal variables within the learned structures, with clear and interpretable causal roles.

Based on these reasonable causal structures, CICDOR further extracts Markov blankets and performs iterative refinement to obtain a more accurate and compact variable set. This refined variable set facilitates more reliable LLM-based confounder extraction, thereby supporting effective deconfounding and robust cross-domain OOD recommendation.

\subsection{Limitations and Future Work}
Our proposed framework focuses on modeling observed confounders that can be explicitly extracted from user reviews. However, unobserved confounders may still exist in real-world recommendation scenarios, such as latent environmental factors. These unobserved confounders can also affect both user preferences and interactions, leading to confounding biases that distort the inferred causal-invariant user preferences. To address this limitation, a promising direction is to extend LLM-guided confounder discovery to the unobserved setting by leveraging the semantic reasoning and world knowledge of LLMs to iteratively hypothesize and validate unobserved confounders \cite{yang2025mitigating}. We leave the systematic modeling of unobserved confounders and corresponding deconfounding strategies as an important direction for future research. In the future, we intend to relax the overlapping-user assumption to further advance the generalizability of our framework. This would allow us to utilize a broader range of cross-domain datasets and explore more diverse and complex OOD scenarios, such as unseen domains and multiple types of distribution shifts.

\section{Conclusion}
In this paper, we have proposed a new setting of cross-domain OOD recommendation, and proposed a novel Causal-Invariant Cross-Domain Out-of-distribution Recommendation framework, called CICDOR. CICDOR consists of a dual-level causal preference learning module to infer domain-specific and domain-shared causal-invariant user preferences, and an LLM-guided confounder discovery module to iteratively identify and refine causal variables from user reviews to extract observed confounders. Through effective deconfounding of the extracted confounders via backdoor adjustment, CICDOR can obtain debiased comprehensive causal-invariant user preferences that significantly improve the OOD recommendation performance in the target domain. Extensive experiments on two real-world datasets validate the superiority of CICDOR over state-of-the-art baselines across various OOD settings.

\begin{acks}
This work is supported by ARC Discovery Project DP230100676.
\end{acks}

\bibliographystyle{ACM-Reference-Format}
\bibliography{reference}

\appendix

\end{document}